# Predictive Modeling: An Optimized and Dynamic Solution Framework for Systematic Value Investing

R.J. Sak*

## Abstract

This paper defines systematic value investing as an empirical optimization problem. Predictive modeling is introduced as a systematic value investing methodology with dynamic and optimization features. A predictive modeling process is demonstrated using financial metrics from Gray & Carlisle and Buffett & Clark. A 31-year portfolio backtest (1985 – 2016) compares performance between predictive models and Gray & Carlisle's Quantitative Value strategy. A 26-year portfolio backtest (1990 – 2016) uses an expanded set of predictor variables to show financial performance improvements. This paper includes secondary novel contributions. Quantitative definitions are provided for Buffett & Clark's value investing metrics. The "Sak ratio" statistic is proposed as an extension to the Benjamini-Hochberg procedure for the inferential identification of false positive observations.

---

* This paper benefited from helpful comments from Ernest P. Chan (QTS Capital Management), Brian L. Betker (Saint Louis University), and Robert A. Korajczyk (Northwestern University). I am grateful to Wesley R. Gray for discussion on implementing the Quantitative Value strategy. All errors and omissions are my own. Thank you Carrie J. Sak, Nancy F. Sak, and Raymond F. Sak for inspiration and support. Manuscript date: September 2017. Address correspondence to: raymondsak2013@u.northwestern.edu

Portions of this work were submitted in thesis from for partial fulfillment of the requirements for the degree of Master of Science in Predictive Analytics from Northwestern University. This research was supported in part through the computational resources and staff contributions provided for the Social Sciences Computing cluster (SSCC) at Northwestern University. Recurring funding for the SSCC is provided by Office of the President, Weinberg College of Arts and Sciences, Kellogg School of Management, the School of Professional Studies, and Northwestern University Information Technology.

**Table of Contents**

## 1. Introduction

Value investing is a stock picking method that exploits relationships between a set of historical financial statement metrics and stock prices. The goal is to pick stocks that consistently produce above average returns, but the problem is that there are infinite ways to define relationships. Systematic value investing is a value investing method that quantitatively defines the relationships between financial statement metrics and stock prices. Quantified relationships enable the use of financial metrics, scoring, and ranking systems for stock picking.

Systematic value investing is an empirical optimization problem. The relationship between a set of historical financial statement metrics and stock prices that produces the best stock return can be algorithmically identified using historical financial statement and stock price data. Predictive modeling is the process of algorithmically identifying the optimal[1] relationship between a set of independent variables[2] and a dependent variable[3]. Predictions about future events are made by applying the relationships to observations with unknown outcomes. This paper introduces predictive modeling as the methodological framework for defining and solving systematic value investing scenarios as optimization problems.

Predictive modeling is a direct descendent of classic value investing. Both assume imperfect relationships[4] between historical financial statement data, stock prices, and future long-term stock price performance. Both exploit these relationships to identify and purchase stocks likely to produce above-average returns. Figure 1 summarizes the steps shared by all value investing strategies. Systematic value investing and predictive modeling use the same steps, but each step may not be explicitly defined. For example, a human analyst might simply describe Step 4 and Step 5 as preferring stocks with relatively higher return on equity than stocks with growing net income. Although a human analyst may not explicitly quantify the relative importance of variables or define how relationships between variables evolve over time, the cognitive tasks are identical to the predictive modeling process.

---

[1] The concept of optimization is relative to the predictive algorithm used. Different algorithms have different optimization methods and definitions of "optimal", but they can generally be described as minimizing the error (or distance) between a predicted outcome and the actual outcome.

[2] Equivalently described in this paper as "financial metrics" and generally described in data science terminology as "predictor variables".

[3] Equivalently described in this paper as "financial outcomes" and generally described in data science terminology as "response variables".

[4] A "perfect relationship" implies a shrewd analysis could identify every stock that produces the desired outcome, with zero errors. Imperfect relationships suggest that intelligent analysis identify a portfolio of stocks that should typically produce results relatively superior to a broad market alternative, such as the S&P 500 Index.

**Figure 1 – Value Investing Steps**

| | |
|---|---|
| Step 1 | Select a desired financial outcome[5] |
| Step 2 | Collect historic financial statement data and stock price data for a universe of companies. |
| Step 3 | Select a set of financial variables to analyze. |
| Step 4 | Select analytical method (criteria) to identify the optimal relationships between the financial variables and the desired financial outcomes. |
| Step 5 | Use the algorithm with the best historical out-of-sample performance to build a stock portfolio using currently available data. |
| Step 6 | Repeat Steps 1-5 as additional data becomes available. |

Defining value investing as an optimization problem implies that predictive modeling must produce results either equal or superior to both classic value investing performed by a human and systematic value investing. A paradox emerges if predictive modeling does not produce superior long-term results. If the analytical methods are described quantitatively or logically, then predictive modeling can at least match, or possibly improve upon (optimize), the results. On the other hand, if the analytical methods cannot be described quantitatively or logically, then the "methods" must be either non-reproducible (luck), non-quantifiable (intuition), or non-sustainable (data mining). Divorcing value investing methods from an empirical, quantitative foundation would pose problems to validity of the value investing philosophy.

This paper adds to the publicly available literature on systematic value investing by proposing and demonstrating the utility of systematic value investing strategies generated by predictive models. The process analyzes a universe of publicly available historical financial statements, uses current stock prices to predict stocks likely to achieve a desired outcome, and builds a stock portfolio based on the predictions. Financial backtests are performed on stock portfolios created by the predictive models. The results demonstrate predictive modeling produces statistically unique investment strategies that generate statistically significant long-term monthly returns in excess of Fama-French 3-factor model and Carhart 4-factor model returns. The results also show that the dynamic and optimized features of predictive models produce better results than traditional (non-optimized, non-dynamic) systematic value investing strategies.

The methods presented in this paper are not exhaustive or unnecessarily sophisticated. Simple methods are prioritized over sophisticated methods to maintain a focus on introducing concepts, demonstrating the flexibility of modeling decisions, and highlighting basic performance benchmarks. Although simple methods may sacrifice performance, the results demonstrate that satisfactory outcomes do not require unnecessarily sophisticated methods.

[5] Example: construct a portfolio that maximizes CAGR over a 10-year horizon.

## 2. Background

The stock market is where buyers and sellers trade ownership claims on companies, called stocks. The goal of stock market participants is to buy and sell stocks that consistently generate above average financial returns relative to the overall market return. Graham & Dodd (1934) proposed a "value investing" philosophy could produce above average financial returns. Value investing is the practice of analyzing a company's historic financial statements to estimate its intrinsic value and value investors purchase stocks when a company's stock price is trading at a sufficient discount (known as a "margin of safety") to the estimated intrinsic value. Value investing evolved into various different investment strategies in the decades following Graham & Dodd. Asness, Frazzini, Israel, & Moskowitz (2015) broadly characterize the value investing strategies as either idiosyncratic or systematic investing. Idiosyncratic value investing is analyzing and purchasing individual stocks. Systematic value investing is using a rules based approach to analyze and purchase a portfolio of stocks from a predefined universe.

Kok, Ribando, & Sloan (2017) find that systematic value investing strategies that use simple financial ratios and metrics (such as price-to-book ratios) fail to produce satisfactory results; comprehensive strategies are needed for satisfactory outcomes. Asness, Frazzini, Israel, & Moskowitz; Kok, Ribando, & Sloan; and Lee (2014) find that using multiple metrics derived from historical financial statement data can be effective systematic value investing strategies and these findings support classic value investing literature. Lee (2014) highlights that Graham & Dodd's 1934 book *Security Analysis* included a ten-point checklist for the quantitative analysis of stocks and Asness, Frazzini, Israel, & Moskowitz highlight that Graham's 1973 edition of *The Intelligent Investor* continued to include checklists for systematic value investing nearly 40 years after *Security Analysis* was published[6]. Graham and Dodd's use of checklists suggest that estimating a company's intrinsic value is possible, but requires the use of several metrics.

Gray & Carlisle's (2012) Quantitative Value strategy (QV) is an example of Kok, Ribando, & Sloan's preferred strategy. QV is a publicly available systematic value investing strategy that performs rigorous financial analysis across several financial statement metrics to maximize a portfolio's compound annual growth rate (CAGR) over a long-term horizon. Given a universe of stocks, QV rigorously analyzes each company by integrating several financial statement metrics and stock price metrics from Sloan (1996), Beneish (1999), Campbell, Hilscher, & Szilagyi (2008), Greenblatt (2006), Piotroski (2000), Boudoukh, Michaely, Richardson, & Roberts (2007), Michaely (2007), Richardson, & Roberts (2007), and Gray & Carlisle. A final score is produced for each stock by combining the metrics into a single number. The stocks are then ranked by score and the highest ranked stocks are chosen for portfolio selection.

---

[6] Others have also noted Graham and Dodd's use of checklists.

QV has two shortcomings. The relationships (coefficients) between variables are not optimized and the variable relationships are static, they are not dynamically updated when new information is available.  QV uses variable coefficients set by both Gray & Carlisle and underlying academic sources. The academic sources use variable coefficients optimized for specific accounting events (e.g., financial statement manipulation), which are not necessarily optimal for investment outcomes. The static variable coefficients assume relationships between the financial metrics and the desired outcome does not change across time and business cycles. Predictive modeling improves upon QV by optimizing variable relationships and dynamically updating the relationships when new financial statement data becomes available.

## 3. Methodology

The QV and predictive modeling strategies are conceptually summarized in Appendix 1 and Appendix 2. QV uses historical financial statement and stock price data to calculate a set of financial metrics from the prior *i* periods for each stock in a universe of stocks.  Financial metrics are calculated for each stock and combined into a single score. The stocks are then ranked by score, and a portfolio is built from stocks with the highest ranking scores. The QV scoring and ranking process repeats at user-defined intervals.

Predictive modeling also uses historical financial statement and stock price data for a universe of stocks. Financial metrics (called "predictor variables") are calculated and a desired quantitative financial outcome ("response variable") is selected. A machine learning algorithm[7] calculates the optimal relationship between the predictor variables and the response variable ("model training"). The actual "predictive model" is the equation that quantifies the relationship between predictor variables and the response variable. Predictions are generated by inputting predictor variables into the equation and the result is the predicted value. Stock portfolios are built from stocks predicted by the model to produce the desired response variable.  A predictive model can pick stocks in real-time by applying the model (calculated equation) to current stock prices and financial statement data.

The predictive model equations developed and backtested in this paper are not shown for two reasons. First, this paper is an introduction to predictive modeling as a systematic value investing methodology. Championing a single model is not the intent and is misleading about the flexibility of a predictive modeling framework. Secondly, the modeling process periodically retrains models using a set of model specifications (Appendix 3). The same specifications are used each time a model[8] is retrained, but the result will produce a different set of equations. It is not necessary to show each model after each training instance because the process design is more important than the resultant model(s). This paper completely describes the modeling building process, which enables the models and results to be reproduced[9].

[7] There are many machine learning algorithms to choose from. This paper uses classification trees, logistic regression, and random forest.

[8] This paper refers to a specific modeling specification as "Model XX", however as noted in the next, the specification will produce different models each time they are retrained.

## 4. Data Structure and Preparation

The dataset contains 712,415 monthly observations between 12/31/1977 – 12/31/2016 composed of annual financial statement data, quarterly financial statement data, and stock price data from the CRSP/Compustat Merged Database (CCM). The dataset does not suffer from survivorship bias because the CCM database includes financial statement and stock price data for delisted companies. Each observation includes a company's month-end closing stock price and financial metrics calculated from financial statements and other publicly available information.[10] The dataset avoids look-ahead bias by lagging the historical availability of financial statement data by 180 days[11] and the dataset does not include financial or utilities companies because of the unique nature of their financial statements and regulatory environments. Figure 2 displays additional Compustat parameters applied to the dataset.

**Figure 2**
**Compustat Dataset Parameters**

| | |
|---|---|
| Global Industry Classification Standard Codes | 4010, 4020, 4030, 4040, and 5510 |
| Consolidation Level | Consolidated |
| Industry Format | Industrial |
| Data Format | Standard |
| Population Source | Domestic |
| Currency | USD |

Several financial metrics are calculated using eight consecutive years of historical financial statement data. The first seven years of monthly observations for each company are omitted because financial metric calculations will be biased with insufficient data. Companies with less than eight years of observations are omitted from the dataset for the same reason. In addition, companies with gaps between available financial statements are not treated as consecutive years. In addition to financial statement data, each observation also uses stock price data to calculate financial metrics. All stock prices use the closing price and all prices are adjusted for stock splits and dividend payments. In the event a stock closing price is not available, but bid/ask data is available, the closing price is set to the midpoint between the bid/ask spread. Observations without a closing price or bid/ask data are removed from the dataset. Various financial metrics require a minimum of one year of monthly stock prices. The first 12 months of observations are omitted for each company and companies with less than one year of stock prices are omitted altogether. Companies with gaps between monthly stock prices are also not treated as consecutive observations.

[9] Appendices 5 – 8 contain exhaustive implementation notes for data pre-processing and variable calculation.
[10] When financial statements (either annual or quarterly) from multiple GVKEYs are associated with a single PERMNO in a particular year, the GVKEY with the more recent publish date is used and the older GVKEY(s) are omitted. This ensures a 1:1 relationship between PERMNO and financial statements for each reporting date.
[11] Example: Suppose Company A released their annual financial statements for 1985 on 4/1/86. The 1985 financial statements would not be joined with Company A's stock price until 10/1/1986.

Predictive models require datasets with complete observations. Several approaches are used for handling missing data. Missing data on retained earnings, current assets, current liabilities and adjusted net income are imputed using formulas provided by Casey, Gao, Kirschenheiter, & Pandit (1993). Missing data for research and development costs is set to zero as suggested by Koh & Reeb (2015). Missing data for dividends is also set to zero. For all other financial statement data item, missing data is imputed using the most recently available observation for each company[12]. Observations with missing stock prices are removed from the dataset. Missing stock prices are not imputed because the validity of the financial backtest relies on empirical stock price data for constructing historically transactable prices.

## 5. Modeling Process

All models are binary classification models. The models are trained to predict (yes/no) whether a set of financial metrics associated with a stock price at time $t_0$ produces annualized monthly returns in excess of a defined benchmark[13] (either Fama-French 3-factor model returns or S&P 500 returns) over the corresponding future *n*-year horizon ($t_{0+n}$). The binary classification models assume that all alphas are equally satisfactory and all negative alphas are equally dissatisfactory. Regression models are excluded to for parsimony[14]. Models are not trained for short selling opportunities[15] because the value investing literature is oriented towards long strategies.

Building predictive models is a process with several decisions about response variables, machine learning algorithms, predictor variables, and selection criteria. Due to the combinational possibilities of various model specifications, it is not efficient to evaluate an exhaustive list of every possible model specification. Decision points have been limited to six response variables, three binary classification algorithms, four predictor variable sets, and two stock selection thresholds for 144 unique model combinations. The total number of models evaluated is further limited to 42 by using an iterative decision process.

Predictive models are built for each item within a decision category and the cross-validated predictive performance is assessed for all items within each category. The item with the best predictive performance is kept for each subsequent decision category and the process continues for each decision category. The model building process completes after all categories are evaluated and a final model is produced.[16] The process is repeated for each of the six response variables and the result is six predictive models that have been optimized on a different response variables. Figure 3 summaries the iterative model building sequence; the default values used in the modeling process are underlined.

---

[12] Other methods for imputing missing data were considered (including more computationally intensive and accurate methods such as multivariate imputation by chained equations), but the "most recently available" method is used for sake of simplicity and computational efficiency.

[13] The various response variables are occasionally collectively referred to as "alphas" when descriptions that are more specific are not necessary.

[14] Training regression models requires setting thresholds for desirable alpha values and defining penalty metrics for overestimating alpha.

[15] Short selling a stock is a profitable when the price of a stock decreases over time. In a short selling operation, an investor borrows a stock and sells it to the market at $t_0$ and repurchases the stock at $t_1$and returns it to the original owner. The investor loses money if the stock increases in price between $t_0$ and $t_1$.

**Figure 3**
**Iterative Development of Predictive Models**

<table>
<tr><th>Decision Category</th><th>Item<br>(Default value underlined)</th><th>Data</th><th>Evaluation Criteria</th></tr>
<tr><td>Binary Classification Algorithm</td><td>• Logistic Regression<br>• Classification trees<br>• Random Forest</td><td rowspan="3">Training<br>(10-fold cross-validated)</td><td rowspan="3">Predictive performance</td></tr>
<tr><td>Predictor Variable Sets</td><td>• <u>QV Basic</u><br>• QV Full<br>• Buffettology<br>• QV Full and Buffettology</td></tr>
<tr><td>Stock Selection</td><td>• <u>50% probability</u><br>• > 90% probability</td></tr>
<tr><td>Response Variable</td><td>• <u>1-year Alpha (Fama-French 3-Factor)</u><br>• 5-year Alpha (Fama-French 3-Factor)<br>• 10-year Alpha (Fama-French 3-Factor)<br>• 1-year Alpha (S&P 500)[17]<br>• 5-year Alpha (S&P 500)<br>• 10-year Alpha (S&P 500)</td><td>Test</td><td>Financial performance and statistical comparisons</td></tr>
</table>

After the six predictive models are developed, financial backtests are performed. Each model generates predictions on out-of-sample test observations and stock portfolios are built from the stocks predicted by each model to achieve corresponding response variable. Models are trained annually on June 30 using all available historic data and stocks are also predicted for purchase on June 30. The models are re-trained on a rolling annual basis (June 30) using all available historic observations and each model's specifications[18] remain unchanged during the backtest. The backtest design also incorporates implementation guidelines provided by Gray & Carlisle. Key details are summarized in Figure 4. Only stocks that actively traded on the historic as-of date are available for purchase. Stocks must also have total market equity above the $40^{th}$ percentile of market equity for all stocks traded on the New York Stock Exchange.[19] Stocks with total market equity less than the 40th percentile breakpoint are not available for purchase during the backtest to ensure sufficient liquidity. Portfolios are constructed using a one-year buy and hold strategy and the portfolios are market capitalization (value) weighted.

[16] Although each combination of predictor variables and response variables is a unique optimization problem, models are comparatively grouped and assessed by response variable only because performance against a defined response variable is what ultimately matters for practical purposes.

[17] "*n*-year S&P 500" is a "yes/no" indicator variable for whether a stock produced compounded monthly returns in excess of S&P 500 returns over a n-year time horizon.

[18] Binary classification algorithm, predictor variables, stock selection criteria, response variable

**Figure 4**
**Backtest Implementation Details**

| Topic | Description |
|---|---|
| Portfolio Weights | Market capitalization weighted<br>One-year buy-and-hold |
| Purchase Frequency | Annually (June 30) |
| Model Training Frequency | Annually (June30) |
| Formation Date | June 30 of year *t* |
| Market Capitalization | NYSE 40% breakpoints (French (2016)) |
| Return Data | Prices adjusted for dividends, splits, and corporate actions |
| Management Fee | 1.5% annualized, paid monthly |
| Transaction Costs | 1.0% transacted amount, paid at each purchase. |
| Taxes | None |

Monthly portfolio values are calculated using historic CRSP stock pricing data in the CCM database. Stocks that delist after their purchase date are assigned a zero price [20] until one of the following events occurs. Delisted stock prices are set to CRSP item "DLPRC" on days when over-the-counter prices or bid/ask information is available. DLPRC is used for single days only and is not carried forward. Prices are set to CRSP item "DLAMT" on days when a distribution payment occurs and DLMT amounts are carried forward since DLAMT represents cash payments transferred to the owner. If neither a DLPRC nor DLAMT event occurs for a delisted stock, then the price remains zero for all trading days after delisting.

Monthly portfolio returns are calculated from the monthly portfolio values. All portfolio returns are post-fees and pre-tax. A 1.5% annualized management fee and 1.0% annual transaction cost are applied to portfolio values prior to calculating monthly returns. No tax is assessed on portfolio values to enable standard comparisons across different tax situations. All financial returns are calculated post management fees and transaction costs. The monthly returns produced by each predictive model are assessed using various financial metrics. The financial performance serves as empirical evidence for investigating whether the use of predictive modeling methods to optimize systematic value investing produces investment strategies that deliver sustained above average performance.

[19] Monthly percentiles are calculated using historical point-in-time breakpoints. NYSE market equity breakpoint historical data retrieved from French (2017).

[20] Zero value is assumed due to zero liquidity and the assumption that the empirical value of a stock is limited to an offering price.

Overfitting Predictive Models and Look-Ahead Bias in Financial Backtests

A predictive modeling process must ensure results will persist on out-of-sample predictions. Overfitting is a problem that occurs when predictive model performs well on in sample training data, but performance deteriorates when the model is deployed on out-of-sample data. Overfitting is addressed by splitting a dataset into training observations and test observations. Training observations are used for model building and assessing out-of-sample predictive performance. Models with the best out-of-sample predictive performance are then subject to financial backtests using previously unused out-of-sample test observations. Generating out-of-sample predictions with training data is achieved by using the 10-fold cross-validation technique suggested by James, Witten, Hastie, & Tibshirani (2013).

The dataset is time-series data and splitting the dataset into training and test observations occurs on a rolling basis relative to the model training date. Observations that occur prior to the model training date are labeled "training" and observations that occur on or after the model training date are labeled "test".[21] A minimum requirement of 1,000 observations is imposed on the training data to ensure sufficient data for using the 10-fold cross-validation method. Consequently, the modeling process begins with observations from June 1978 (including trend variables utilizing financial data dating back to June 1971).

A time lag based on the response variable is also added between training observations and test observations to avoid look-ahead bias during the financial backtest. For example, suppose a binary classification model is trained on 6/30/90 to predict whether a stock will produce a 5-year Fama-French 3-factor alpha. The final training observation must be dated 6/30/85 because five years have not elapsed for observations between 7/1/85 and 6/29/30 and it is unknown whether these observations achieved the desired five-year alpha. The time lag for each model iteration is dependent on the response variable (e.g., ten-year response variable requires a ten-year lag). See Appendix 2 for a conceptual diagram.

## 6. Model Performance Evaluation

The predictive modeling process is performed in two steps. First, models are iteratively developed and out of sample predictive performance is assessed. This process occurs for six different response variables, which produces six unique models. Second, financial backtests are performed by using each of the six models to build a stock portfolio based on each model's binary stock predictions. Financial performance is measured by assessing the monthly returns of each stock portfolio.

[21] For example, given a 6/30/90 model training date, predictive models are built and evaluated for predictive performance using observations prior to 6/30/90. Financial performance is then assessed by using the model to make predictions on observations starting with 6/30/90. All models are retrained annually (See Figure 6) and the process repeats on 6/30/91.

Predictive Performance Metrics

The predictive models are binary classification models, they simply predict (yes/no) whether a stock trading at price *p* on day $t_o$ will achieve a predefined financial outcome[22] over a predefined time period $t_{0+n}$, given a set of financial metrics. Predictive performance is assessed using three classification metrics: precision, the false negative rate, and area under the receiver operating characteristic curve (AUC). All predictive metrics are calculated from a confusion matrix (Figure 5), a 2×2 contingency table that compares the predicted outcome of a binary classification problem with the actual outcome.

The precision and false negative rate metrics are defined in Figure 6. Precision measures the percentage of stocks predicted to achieve the desired response variable that actually accomplish it. A high precision statistic is good; a low precision statistic will result in investing in stocks that fail to deliver the desired benchmark. The false negative rate measures the percentage of stocks not predicted to beat the alpha benchmark that actually end up producing alpha. The false negative rate implies opportunity cost and a low false negative rate is preferred. The AUC metric measures whether a binary classification model's performance can be distinguished from random selection after accounting for the underlying distribution of event outcomes. AUC scores range between 50% (indistinguishable from random chance) and 100% (pure signal); scores above 80% are preferred[23].

**Figure 5**
**Confusion Matrix**

| Actual Outcome | Predicted Outcome: Alpha | Predicted Outcome: Not Alpha |
|---|---|---|
| Alpha | True Positive (TP) | False Negative (FN) |
| Not Alpha | False Positive (FP) | True Negative (TN) |

**Figure 6**
**Confusion Matrix Metrics**

$$Precision = \frac{TP}{(TP + FP)}$$

$$False\ Negative\ Rate = \frac{FN}{(FN + FP)}$$

$$AUC = See\ Hanley\ \&\ McNeil$$

---

[22] See Appendix 3 for an overview of the various response variables.
[23] See Hanley & McNeil (1982) for a technical AUC definition.

## Financial Performance Metrics

Financial metrics for assessing backtest performance are sourced from Gray & Carlisle. The metrics are described in Figure 7. Using several metrics ensures a robust assessment of financial performance. The multi-factor model metrics are the primary performance measures. Multi-factor models describe stock returns in terms of financial and market characteristics (factors). Two common multi-factor models are the Fama-French 3-factor model (Fama & French, 1993) and the Carhart 4-factor model (Carhart, 1997). The Fama-French 3-factor model describes stock returns as a function of a market risk factor, a value factor, and a size factor. The Carhart 4-factor model adds a fourth momentum factor. Assessing whether an investment strategy produces monthly returns in excess of a multi-factor model is an important test. Failing to generate excess returns suggests that an investment strategy does not add value because a passive investment strategy could theoretically achieve similar returns.

**Figure 7 – Financial Metrics for Assessing Backtest Performance**

| Metric | | Description |
|---|---|---|
| CAGR | | Compound annual growth rate (CAGR) of monthly returns for 1990 – 2016. |
| Standard deviation | | Standard deviation of all monthly returns (annualized by $\sqrt{12}$). |
| Downside deviation | | Standard deviation of negative monthly returns (annualized by $\sqrt{12}$). |
| Information ratio | | Mean of monthly returns minus monthly S&P 500 returns divided by standard deviation of monthly return minus monthly S&P 500 returns (annualized by $\sqrt{12}$). |
| Sortino ratio | | Monthly return minus minimum acceptable return (5%/12) divided by downside deviation (annualized by $\sqrt{12}$). |
| Profitable months | | Percentage of monthly returns > 0. |
| Best return | | Best monthly portfolio return. |
| Worst return | | Worst monthly portfolio return. |
| Worst drawdown | | Worst peak-to-trough performance of portfolio value. |
| Cumulative drawdown | | Sum of the worst drawdowns over rolling monthly five-year windows. |
| FF 3-factor model | Alpha (annualized) | CAGR of monthly returns regressed against the Fama-French 3-factor model for 1990 – 2016 (annualized figure). |
| | Alpha (p-value) | Probability value associated with the alpha figure. |
| | Alpha (adjusted $R^2$) | Adjusted $R^2$ for monthly returns regressed against the Fama-French 3-factor model. |
| | Alpha (f-statistic) | F-statistic for monthly returns regressed against the Fama-French 3-factor model. |
| | 1-year alpha windows | Percentage of rolling monthly one-year windows with alpha > 0. |
| | 5-year alpha windows | Percentage of rolling monthly five-year windows with alpha > 0. |
| | 10-year alpha windows | Percentage of rolling monthly ten-year windows with alpha > 0. |
| Carhart 4-factor model | Alpha (annualized) | CAGR of monthly returns regressed against the Carhart 4-factor model for 1990 – 2016 (annualized figure). |
| | Alpha (p-value) | Probability value associated with the alpha figure. |
| | Alpha (adjusted $R^2$) | Adjusted $R^2$ for monthly returns regressed against the Carhart 4-factor model. |
| | Alpha (f-statistic) | F-statistic for monthly returns regressed against the Carhart 4-factor model. |
| | 1-year alpha windows | Percentage of rolling monthly one-year windows with alpha > 0. |
| | 5-year alpha windows | Percentage of rolling monthly five-year windows with alpha > 0. |
| | 10-year alpha windows | Percentage of rolling monthly ten-year windows with alpha > 0. |

Statistical Controls for Interpreting Financial Performance Metrics

Identifying backtested statistically significant monthly portfolio returns suffers from the multiple comparisons problem. Performing multiple backtests on the same test observations increases the likelihood that a traditional 0.05 p-value threshold will falsely identify statistically significant results. Failing to correct the multiple comparisons problem may result in accepting results that are data mined and not reproducible. The multiple comparisons problem is addressed in the results section by using the Benjamini-Hochberg procedure (Benjamini & Hochberg, 1995), which identifies statistical significance by controlling the false discovery rate[24], instead of adjusting the p-value significance threshold. McDonald (2009) favors the Benjamini-Hochberg procedures over the more conservative Bonferroni Correction[25] for scenarios where an acceptable false discovery rate is greater than zero.

The Wilcoxon signed-rank test (Wilcoxon, 1945) assesses whether the predictive models described in this paper represent different investment strategies. The distribution of monthly Fama-French 3-factor and Carhart 4-factor alphas generated by each model are compared based on the assumption that different investment strategies provide different factor-adjusted returns. The Wilcoxon signed-rank test is used instead of the Student's t-test. Officer (1972) finds that stock market returns do not follow a normal distribution and the Wilcoxon tests do not assume data has a normal distribution. Additionally, Wilcoxon signed-rank tests are performed on both Fama-French 3-factor model and Carhart 4-factor model alphas to err on the side of caution.

## 7. Predictive Performance Results

The following section provides the results from iteratively developing predictive models using six different response variables. As shown previously in Figure 3, considering three binary classification algorithms, four predictor variable sets, and two stock selection thresholds produce 24 unique predictive models per response variable (144 models total). The iterative approach reduces the number of models for consideration to seven per response variable (42 models total). Although the iterative approach is not exhaustive and may not identify the model capable of producing the best results, it is an efficient and reasonable process for identifying models with good results.

For sake of brevity, full results for all 42 models are not displayed. The following section shows full predictive performance results for all seven models using the one-year Fama-French 3-factor model alpha response variable to demonstrate the process for iterative development and performance assessment. The same process was performed using five other response variables (see Figure 3 for details), but results from the 35 other models are not shown for sake of brevity.

---

[24] The false discovery rate is the percentage of statistically significant outcomes that are actually false positives (not statistically significant).

[25] Instead of rejecting $H_0$ when p-value$_t$ $< 0.05$, the Bonferroni Correction sets the critical threshold $< \frac{0.05}{n}$, where $n$ = total comparisons.

Predictive performance is assessed by using a 10-fold cross validation to train models and produce out-of-sample predictions (See Appendix 2 for details). The model training/validation process is performed monthly between 1981 – 2015 for 420 sets of out-of-sample predictions. Predictive performance metrics are calculated for each set of predictions and the results are plotted as a time series to assess the durability of each model[26].

Predictive Algorithms

The model building process begins with considering three different algorithms for binary classification predictions: logistic regression (Model 1), classification tree (Model 2), and random forest (Model 3). See Appendix 4 for a brief conceptual introduction to each predictive algorithm. All algorithms are implemented using the statistical programming language R. The logistic regression algorithm uses the `glm` package, the classification tree algorithm uses the `rpart` package, and the random forest algorithm uses the `randomForest` package. Default tuning parameters are used for each algorithm, except the `randomForest` *ntree* parameter is set to 21[27]. More sophisticated algorithms, ensembling, and/or algorithm tuning could improve performance, however implementing relatively, unturned algorithms promotes parsimony and eliminates questions about whether excessively complex predictive modeling methods are necessary for satisfactory performance.

10-fold cross-validated results are presented in Figure 8. The random forest algorithm (Model 3) performs best across all metrics and it will be used for all subsequent iterations. Model 3 predictive performance is stable across 35 years of monthly predictions which suggests the random forest algorithm is reliably using financial statement data and stock pricing data to identify stocks that will produce Fama-French 3-factor alpha over a one-year horizon.

**Figure 8 – Cross-Validated Results Across Different Predictive Algorithms**

Model 1 Model 2 Model 3

**AUC**

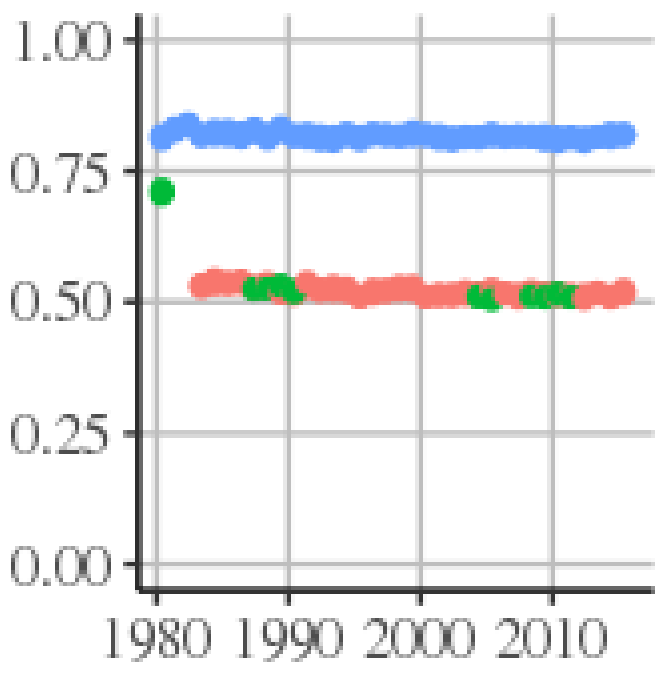

| Model | Std. Dev. | Min. | Median | Mean | Max |
|---|---|---|---|---|---|
| Model 1 | 0.007 | 0.510 | 0.517 | 0.520 | 0.535 |
| Model 2 | 0.057 | 0.509 | 0.516 | 0.535 | 0.705 |
| Model 3 | 0.005 | 0.809 | 0.817 | 0.818 | 0.837 |

[26] The predictive modeling specifications applied to each data. See Appendix 3 for details.
[27] *ntree* parameter set to 21and parallelized across three computing cores

**Precision**

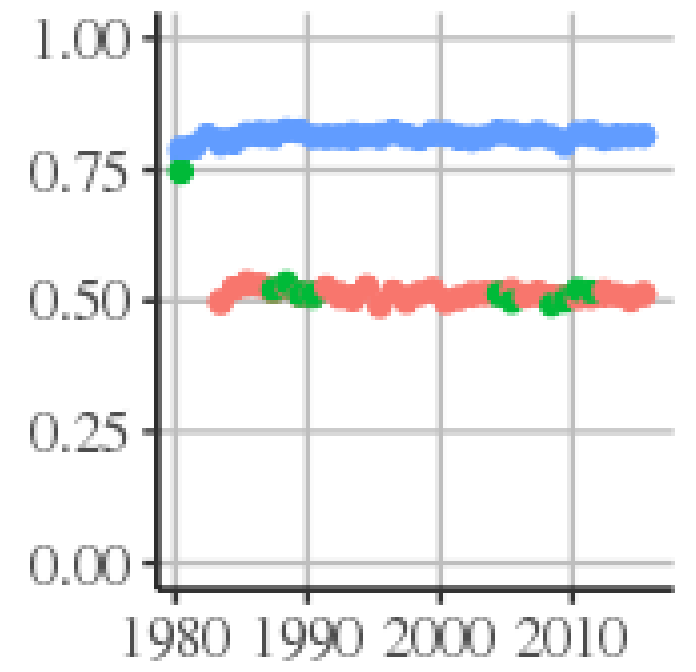

| Model | Std. Dev. | Min. | Median | Mean | Max |
|---|---|---|---|---|---|
| Model 1 | 0.008 | 0.494 | 0.511 | 0.513 | 0.531 |
| Model 2 | 0.072 | 0.501 | 0.513 | 0.534 | 0.750 |
| Model 3 | 0.007 | 0.783 | 0.813 | 0.812 | 0.823 |

**False Negative Rate**

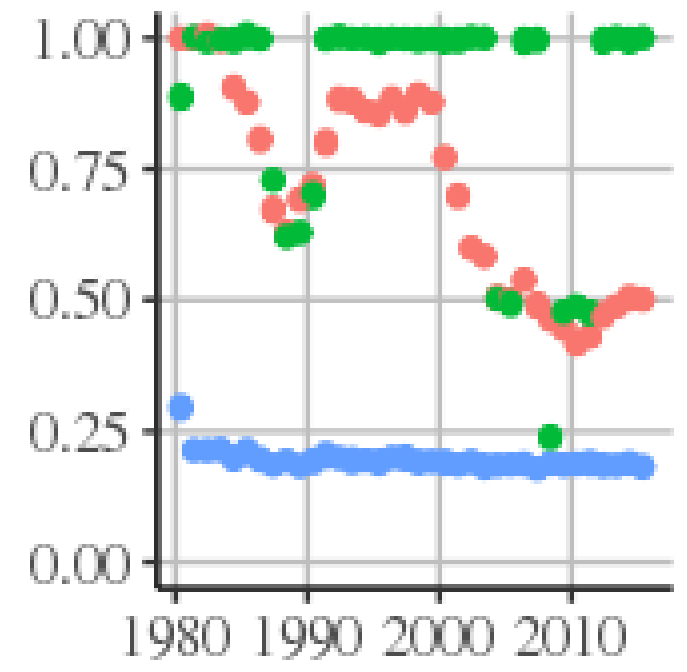

| Model | Std. Dev. | Min. | Median | Mean | Max |
|---|---|---|---|---|---|
| Model 1 | 0.195 | 0.416 | 0.711 | 0.709 | 1.000 |
| Model 2 | 0.222 | 0.236 | 1.000 | 0.867 | 1.000 |
| Model 3 | 0.020 | 0.180 | 0.189 | 0.196 | 0.300 |

<u>Predictor Variables</u>

Predictor variables from two sources are assessed: Gray & Carlisle and Buffett & Clark's 1997 book *Buffettology*. Buffett & Clark is a primary source for famed value investor Warren Buffett's investing metrics and methods. Mary Buffett is Warren Buffett's former daughter-in-law and Buffett & Clark is her interpretation of conversations with Warren about his investing methods. Buffett & Clark provide several metrics for assessing a company's financial performance and intrinsic value. Buffett & Clark also organize the metrics into an investing checklist. One shortcoming of the Buffett & Clark metrics is the use of subjective language (e.g., "strong earnings", "upward trends", etc.) which prevents their use in systematic value investing strategies (including predictive models). This paper makes a novel contribution to Buffet & Clark by defining the variables in purely quantitative terms, which enables their use in quantitative and algorithmic investment strategies. Another shortcoming is Buffett & Clark do not prioritize their checklist. Predictive modeling solves this shortcoming by algorithmically identifying the optimal relationships (relative priority) between the checklist variables. See Appendices 7 and 8 for more information.

Metrics from Gray & Carlisle's QV strategy are mainly sourced from academic finance and accounting papers.[28] The various papers model and predict events that investors find interesting

[28] See Appendix 4 and Appendix 5 for details.

(e.g., financial statement manipulation[29]). Predictive modeling also discovers the optimal relationships between the Gray & Carlisle variables, which reduces dependence on potentially suboptimal coefficients set by the academic source. See Appendices 5 and 6 for details on both QV variable sets.

Four sets of predictor variables are assessed: QV Basic (Model 3), QV Full (Model 4), Buffettology (Model 5), and QV Full plus Buffett & Clark variables (Model 6). QV Basic comprises the final five variables used in Gray & Carlisle's Quantitative Value strategy and QV Full includes 53 constituent variables used in calculating the QV Basic variables. The Buffett & Clark variables comprise 18 variables used for quantifying Buffett & Clark's checklist. The QV Full plus Buffett & Clark variable set combines all variables from the aforementioned set.

Figure 9 shows the 10-fold cross-validated results for each predictor variable set. The QV Full plus Buffett & Clark variable set performs best across all metrics. Performance is similar between the other three variable sets, and performance appears to generally improve as more variables are included in the model. In addition, performance is relatively stable across time for all four models.

**Figure 9 – Cross-Validated Results Across Different Response Variable Sets**

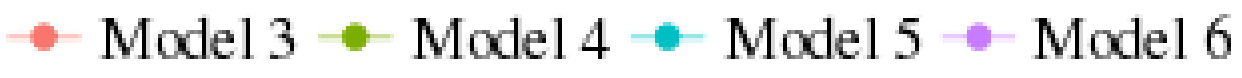

**AUC**

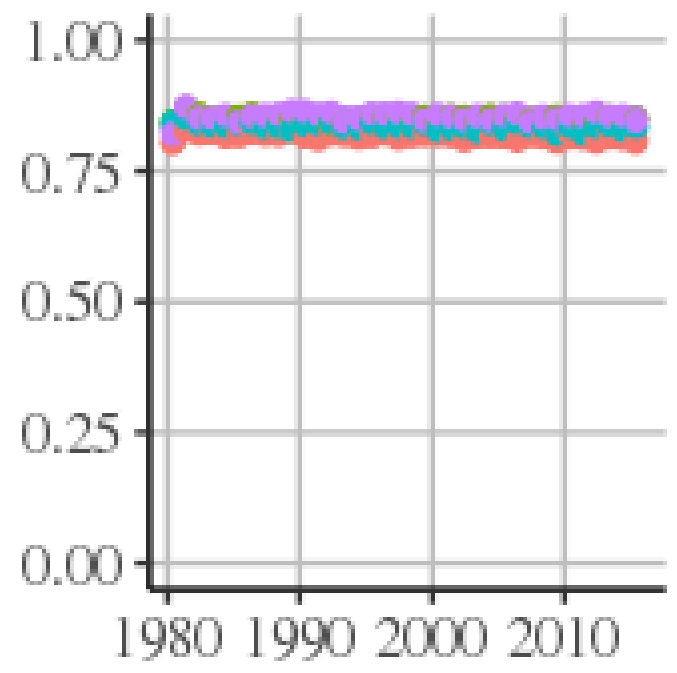

| Model | Std. Dev. | Min. | Median | Mean | Max |
|---|---|---|---|---|---|
| Model 3 | 0.006 | 0.805 | 0.816 | 0.817 | 0.836 |
| Model 4 | 0.003 | 0.845 | 0.851 | 0.851 | 0.862 |
| Model 5 | 0.004 | 0.829 | 0.834 | 0.834 | 0.851 |
| Model 6 | 0.006 | 0.831 | 0.854 | 0.854 | 0.872 |

**Precision**

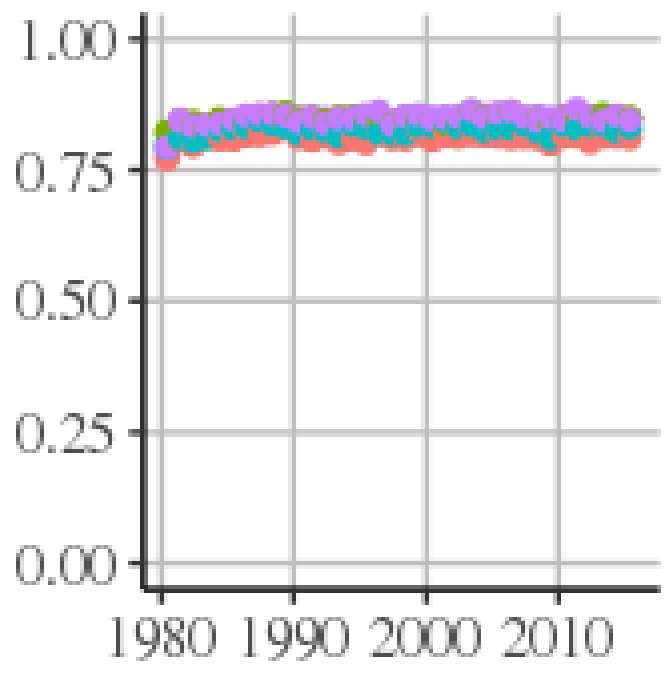

| Model | Std. Dev. | Min. | Median | Mean | Max |
|---|---|---|---|---|---|
| Model 3 | 0.008 | 0.778 | 0.812 | 0.812 | 0.829 |
| Model 4 | 0.007 | 0.820 | 0.847 | 0.846 | 0.856 |
| Model 5 | 0.008 | 0.801 | 0.830 | 0.827 | 0.841 |
| Model 6 | 0.012 | 0.790 | 0.851 | 0.848 | 0.858 |

[29] For example, see the $pman_t$ variable in Appendix 5, Checklist Item 1.1.2

| False Negative Rate | |
|---|---|
| 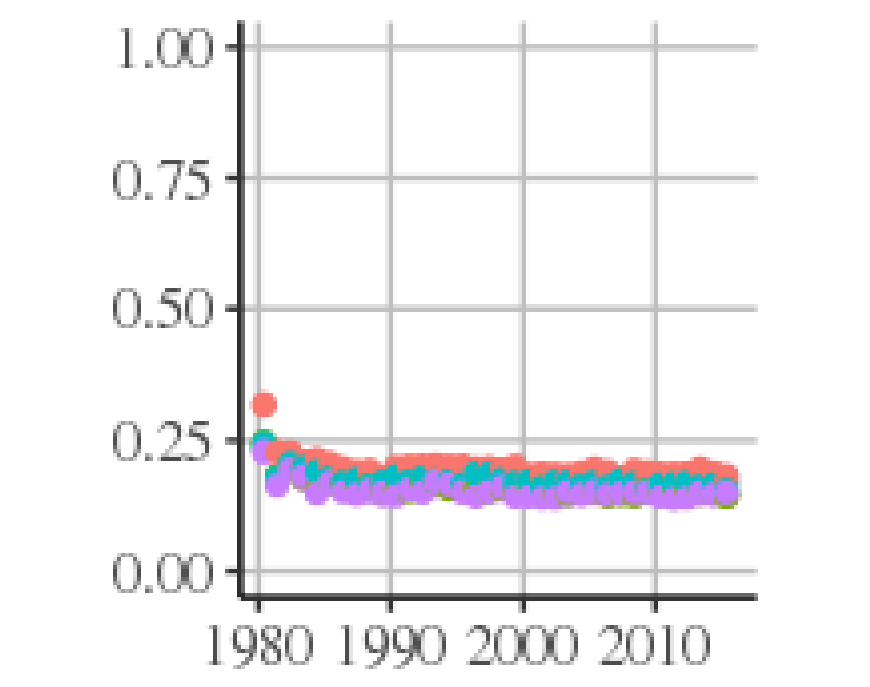 | |

| Model | Std. Dev. | Min. | Median | Mean | Max |
|---|---|---|---|---|---|
| Model 3 | 0.023 | 0.180 | 0.193 | 0.197 | 0.317 |
| Model 4 | 0.017 | 0.145 | 0.155 | 0.159 | 0.239 |
| Model 5 | 0.015 | 0.163 | 0.172 | 0.176 | 0.239 |
| Model 6 | 0.017 | 0.144 | 0.151 | 0.156 | 0.228 |

<u>Stock Selection Cut-off Threshold</u>

Stock selection refers to adjusting the cutoff threshold for binary classification models. The random forest algorithm outputs a predicted probability that a particular stock will produce a Fama-French 3-factor alpha over a one-year return. The default cutoff threshold for classification algorithms is 50%; if the predicted probability is >50%, then a stock is predicted to return alpha. The "stock selection" category tests whether predictive performance can be improved by using a different cutoff threshold for stock selection. Two thresholds are considered: f >50% predicted probability (Model 6) and the ninth decile (top 10%) of predicted probabilities across all observations on a cumulative rolling basis (Model 7). The ninth decile is subjectively chosen because it represents the top 10% of observations with the highest predicted probability for alpha, which is different from simply selecting all observations with >50% predicted probability for alpha.

10-fold cross-validated results are presented in Figure 10. Model 6 performs better than Model 7 for the AUC and false negative rate metrics, but Model 7 performs better on precision. Overall, Model 6 performs best because the AUC metric is prioritized for assessing predictive performance. In addition, as with previous iterations, both models have relatively stable performance across time periods.

**Figure 10 – Cross-Validated Results Across Different Selection Cutoff Thresholds**

## AUC

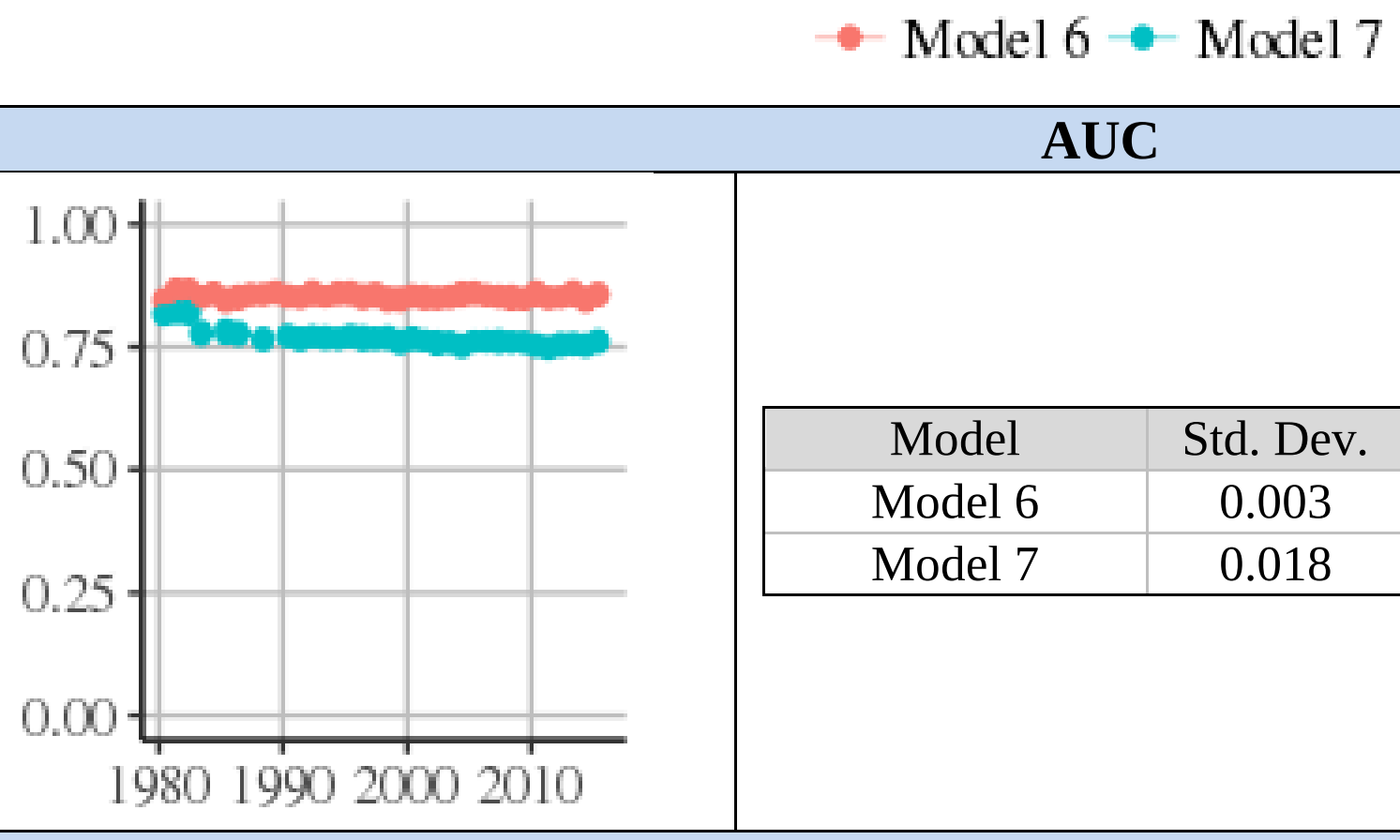

| Model | Std. Dev. | Min. | Median | Mean | Max |
|---|---|---|---|---|---|
| Model 6 | 0.003 | 0.849 | 0.854 | 0.854 | 0.863 |
| Model 7 | 0.018 | 0.752 | 0.764 | 0.770 | 0.825 |

## Precision

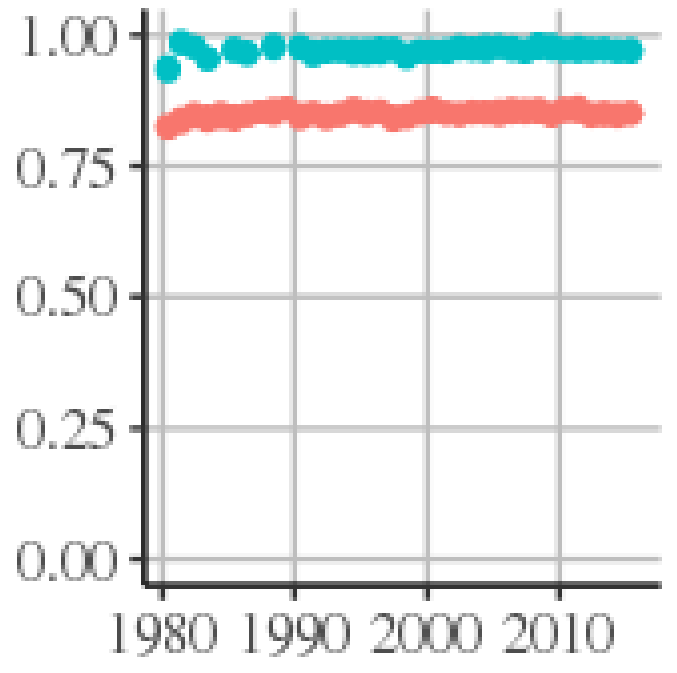

| Model | Std. Dev. | Min. | Median | Mean | Max |
|---|---|---|---|---|---|
| Model 6 | 0.007 | 0.822 | 0.850 | 0.849 | 0.860 |
| Model 7 | 0.008 | 0.940 | 0.972 | 0.971 | 0.988 |

## False Negative Rate

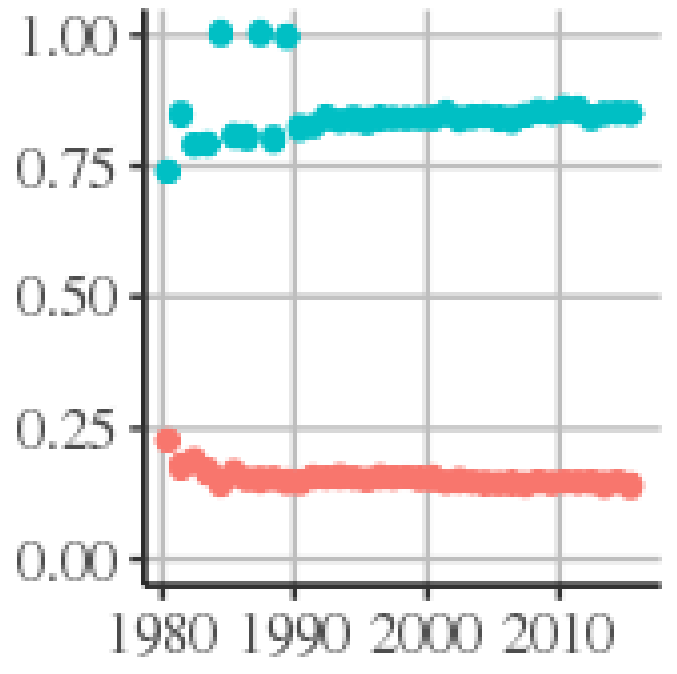

| Model | Std. Dev. | Min. | Median | Mean | Max |
|---|---|---|---|---|---|
| Model 6 | 0.016 | 0.144 | 0.150 | 0.156 | 0.228 |
| Model 7 | 0.052 | 0.739 | 0.843 | 0.848 | 1.000 |

<u>Summary of Predictive Results Across Other Response Variables</u>

The first three iterative categories considered different predictive algorithms, predictor variable sets, and selection cut-off thresholds. The same binary response variable (1-year Fama-French 3-factor alpha) was used for all models. Predictive performance was assessed across 1981 – 2015 by measuring AUC, precision, and false negative rate metrics on out-of-sample predictions made using a 10-fold cross-validation on training data. The best results were produced by a random forest algorithm set to a 50% cutoff threshold using a combined predictor set of QV Full and Buffettology predictor variables. In addition, the monthly performance produced by the random forest algorithm across all model iterations showed relatively low variance across the 35-year horizon.

The preceding process was repeated five additional times using different response variables: Fama-French five-year alpha (Model 8), Fama-French ten-year alpha (Model 9), one-year S&P 500 alpha (Model 10), five-year S&P 500 alpha (Model 11), and ten-year S&P 500 alpha (Model 12). Predictive performance was evaluated using the same criteria described in previous sections and all five models converged to the same modeling parameters as Model 6. A random forest algorithm trained on the QV Full and Buffett & Clark predictor variables with stock selection set to >50% threshold produced optimized predictive performance across all six models.

## 8. Financial Performance Results

Financial backtests were performed on the best predictive models from each of six response variables. The results are shown in Figure 11 and the results are limited to June 1990 – December 2016 to ensure a consistent time horizon for comparison between 1-year, 5-year, and 10-year response variables[30]. Overall, models trained on S&P 500 alpha response variables (Models 10 – 12) performed better than the models trained on Fama-French 3-factor model response variables (Models 6, 8, 9) across nearly every financial metric. Figure 11 also provides statistics for each model's ability to produce alpha on a rolling monthly basis (*x*-year excess return periods). The figures are the percentage of months with a multi-factor model alpha > 0 over 1-year, 5-year, and 10-year horizons. Model 11 and Model 12 both produced 10-year positive alphas for both multi-factor models across 100% of months.

[30] Backtest start dates vary due to varying lag between training observations and test observations caused by response variables with different time horizons. For example, Models 6 and 10 begin in 1981, Models 8 and 11 begin in 1985, and Models 9 and 12 begin in 1990.

## Figure 11 – Backtest Results Across Different Response Variables
## June 1990 – December 2016

| | | Model 6 | Model 8 | Model 9 | Model 10 | Model 11 | Model 12 |
|---|---|---|---|---|---|---|---|
| Response variable | | Fama-French 1-year alpha | Fama-French 5-year alpha | Fama-French 10-year alpha | S&P 500 1-year alpha | S&P 500 5-year alpha | S&P 500 10-year alpha |
| CAGR | | 11.52% | 12.33% | 10.72% | 13.51% | 15.92% | 17.61% |
| Standard deviation | | 17.77% | 18.52% | 17.13% | 17.33% | 17.71% | 23.08% |
| Downside deviation | | 3.96% | 4.08% | 3.84% | 3.85% | 3.82% | 4.40% |
| Information ratio | | 0.350 | 0.419 | 0.600 | 0.570 | 0.760 | 0.590 |
| Sortino ratio | | 0.189 | 0.202 | 0.168 | 0.236 | 0.289 | 0.302 |
| Profitable months | | 57.99% | 58.31% | 57.68% | 60.19% | 61.13% | 59.25% |
| Best return | | 18.86% | 19.68% | 15.38% | 18.52% | 21.62% | 45.17% |
| Worst return | | -23.78% | -25.06% | -23.40% | -23.97% | -23.95% | -22.99% |
| Worst drawdown | | -52.56% | -53.48% | -51.73% | -48.66% | -47.96% | -48.05% |
| Cumulative drawdown | | -319,888% | -323,907% | -310,022% | -284,261% | -275,104% | -309,420% |
| FF 3-factor model | Alpha (annualized) | 0.77% | 1.80% | 0.66% | 2.85% | 5.51% | 7.34% |
| | Alpha (p-value) | 0.552 | 0.183 | 0.597 | 0.045 | 0.001 | 0.022 |
| | Alpha (adjusted $R^2$) | 0.863 | 0.864 | 0.866 | 0.828 | 0.764 | 0.508 |
| | Alpha (f-statistic) | 0.000 | 0.000 | 0.000 | 0.000 | 0.000 | 0.000 |
| | 1-year alpha windows | 59.74% | 58.44% | 58.12% | 62.66% | 68.51% | 71.43% |
| | 5-year alpha windows | 59.23% | 78.46% | 67.69% | 80.00% | 91.15% | 92.69% |
| | 10-year alpha windows | 57.50% | 100.00% | 59.00% | 92.00% | 100.00% | 100.00% |
| Carhart 4-factor model | Alpha (annualized) | 1.80% | 2.72% | 1.62% | 3.82% | 6.41% | 8.01% |
| | Alpha (p-value) | 0.161 | 0.043 | 0.186 | 0.007 | 0.000 | 0.014 |
| | Alpha (adjusted $R^2$) | 0.871 | 0.870 | 0.874 | 0.836 | 0.770 | 0.509 |
| | Alpha (f-statistic) | 0.000 | 0.000 | 0.000 | 0.000 | 0.000 | 0.000 |
| | 1-year alpha windows | 55.84% | 56.82% | 60.06% | 60.39% | 70.45% | 66.88% |
| | 5-year alpha windows | 79.23% | 80.38% | 77.31% | 88.08% | 92.69% | 92.69% |
| | 10-year alpha windows | 98.00% | 100.00% | 97.50% | 100.00% | 100.00% | 100.00% |

Figure 12 charts the rolling alpha figures and it also charts the relative value of all six portfolios relative to the S&P 500 index for the 1990 – 2016 backtest period. The rolling alpha figures for both models consistently improved as the time horizon increased from 1-year to 10-year, which aligns with traditional value investing theory that value investing is most suitable for long term capital appreciation.

**Figure 12 – Rolling Alpha Comparison**
**June 1990 – December 2016**

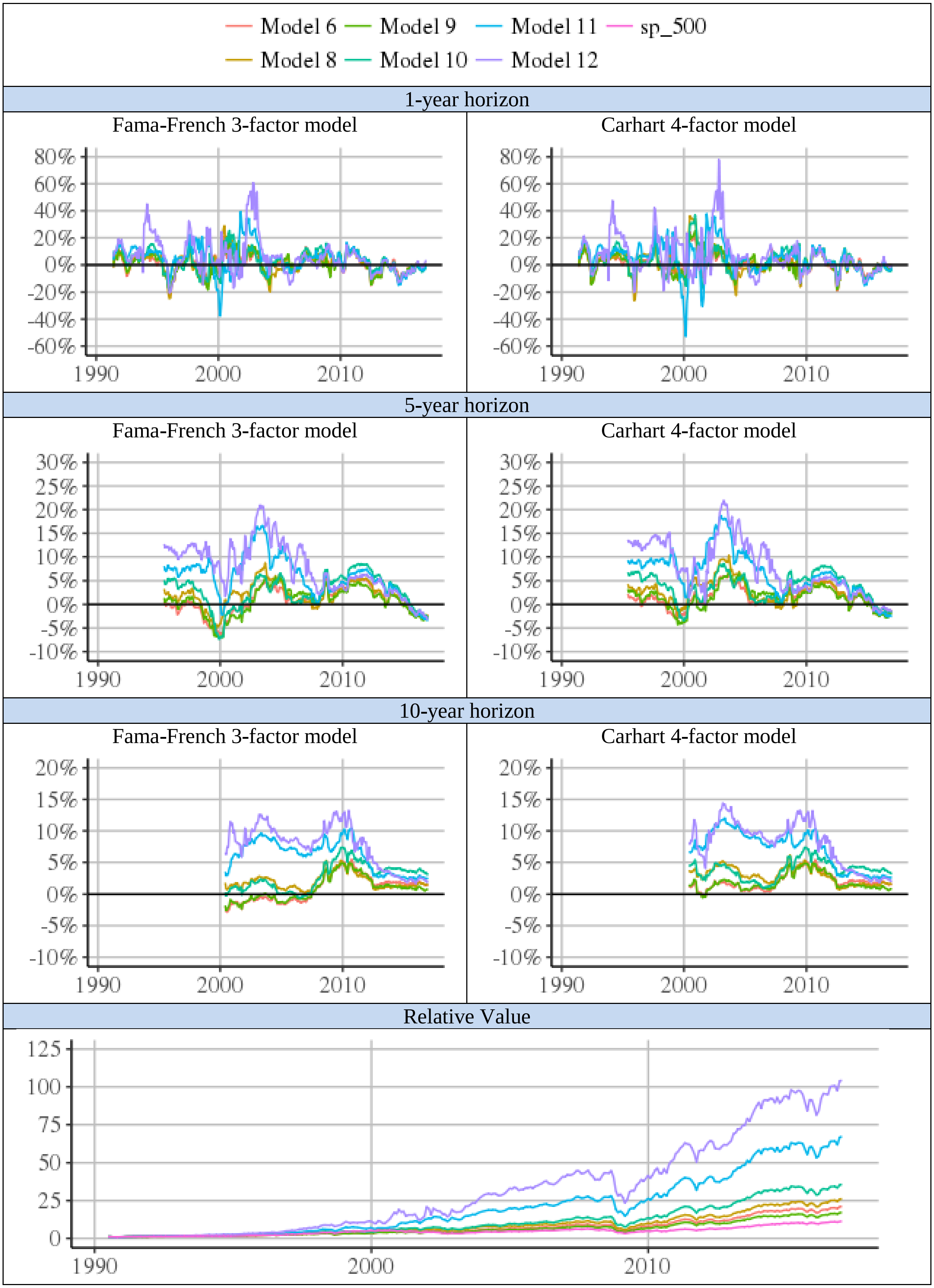

The results of applying the Benjamini-Hochberg procedure to the alpha figures are shown in Figure 13. A 0.095 q-value[31] implies that three models (8, 10-12) produced backtested monthly returns with statistically significant 3-factor alpha and four models (8, 10-12) produced statistically significant 4-factor alpha monthly returns. The 0.095 q-value implies 0.29 false positives for the four 3-factor alpha models and 0.38 false positives for the three 4-factor alpha models. The Benjamini-Hochberg procedure cannot infer the distribution probability of false negative counts nor can it assess which Models are more likely to be among the false positives. More work is needed to identify models most likely to be false positives.

**Figure 13 – Benjamini-Hochberg Procedure[32]**
**q-value = 0.095**

| Model | Fama-French 3-Factor Alpha | | | | Carhart 4-Factor Alpha | | | |
|---|---|---|---|---|---|---|---|---|
| | p-value | Rank | Threshold | Significant? | p-value | Rank | Threshold | Significant? |
| Model 6 | 0.552 | 5 | 0.079 | No | 0.161 | 5 | 0.079 | No |
| Model 8 | 0.183 | 4 | 0.063 | No | 0.043 | 4 | 0.063 | Yes |
| Model 9 | 0.597 | 6 | 0.095 | No | 0.186 | 6 | 0.095 | No |
| Model 10 | 0.045 | 3 | 0.048 | Yes | 0.007 | 2 | 0.032 | Yes |
| Model 11 | 0.001 | 1 | 0.016 | Yes | 0.000 | 1 | 0.016 | Yes |
| Model 12 | 0.022 | 2 | 0.032 | Yes | 0.014 | 3 | 0.048 | Yes |
| | | | Total | 3 | | | Total | 4 |
| | Expected false positives (q-value = 0.095) | | | 0.29 | Expected false positives | | | 0.38 |

Calculating the cumulative distribution function (CDF) for the Benjamini-Hochberg procedure results provides probability estimates for total false positive counts. Figure 14 shows the probability of total false positives for both the Fama-French Alpha and the Carhart Alpha. The CDF uses a 0.095 event success rate (q-value) and six trials (one for each financial backtest). The results in Figure 14 do not suffer from the multiple comparisons problem, so a standard 0.05 probability threshold can be used to identify false negative counts. Figure 14 infers that one of the significant Fama-French returns (Figure 13) is a false positive and one of the significant Carhart returns is also a false positive.

**Figure 14 – Using the Cumulative Distribution Function**
**To Infer False Positive Counts**

| Fama-French 3-Factor Alpha | |
|---|---|
| False Positives | Probability |
| > 0 | 0.259 |
| > 1 | 0.025 |
| > 2 | 0.001 |
| > 3 | 0.000 |
| > 4 | -- |
| > 5 | -- |
| 6 | -- |

| Carhart 4-Factor Alpha | |
|---|---|
| False Positives | Probability |
| > 0 | 0.329 |
| > 1 | 0.048 |
| > 2 | 0.003 |
| > 3 | 0.000 |
| > 4 | 0.000 |
| > 5 | -- |
| 6 | -- |

[31] The q-value is the maximum acceptable false positive rate. The Benjamini-Hochberg procedure ensures the realized false positive rate is less than or equal to the desired q-value.
[32] Data shown for reproducibility See Benjamini & Hochberg and McDonald for calculation steps.

The "Sak ratio" is the ratio between the Benjamini-Hochberg threshold and the p-value. It can be used to infer which of the alphas in Figure 13 are most likely false positives. Relatively lower significant ratios indicate relatively higher false positive probability. Since Figure 14 inferred one false positive for both the Fama-French and Carhart alphas, the Sak ratios in Figure 15 suggest that the 3-factor alpha statistic for Model 10 and the 4-factor alpha statistic for Model 8 are both false positives (not statistically significant). In addition, these results are inferential evidence suggesting the monthly returns for Model 11 and Model 12 produced statistically significant Fama-French 3-factor positive alpha and Carhart 4-factor positive alpha over a 26-year backtest. [33]

**Figure 15 –Using Sak Ratio to Inferentially Identify False Positives**

| Model | Fama-French 3-Factor Alpha | | | | Carhart 4-Factor Alpha | | | |
|---|---|---|---|---|---|---|---|---|
| | p-value | Threshold | Sak Ratio | Likely Significant? | p-value | Threshold | Sak Ratio | Likely Significant? |
| Model 6 | 0.552 | 0.079 | -- | No | 0.161 | 0.079 | -- | No |
| Model 8 | 0.183 | 0.063 | -- | No | 0.043 | 0.063 | 1.48 | False Positive |
| Model 9 | 0.597 | 0.095 | -- | No | 0.186 | 0.095 | -- | No |
| Model 10 | 0.045 | 0.048 | 1.04 | False Positive | 0.007 | 0.032 | 4.43 | Yes |
| Model 11 | 0.001 | 0.016 | 35.40 | Yes | 0.000 | 0.016 | 79.44 | Yes |
| Model 12 | 0.022 | 0.032 | 2.02 | Yes | 0.014 | 0.048 | 3.34 | Yes |

Next, the backtested monthly returns from Model 11 and Model 12 are tested for statistically significant differences using the Wilcoxon signed-rank test. The Wilcoxon signed-rank test assesses whether Model 11 and Model 12 represent different investment strategies because different investment strategies are assumed to produce different monthly returns, The Wilcoxon signed-rank test compares rolling monthly windows for 3-factor and 4-factor alpha returns over 1-year, 5-year, and 10-year horizons for both models. The Wilcoxon signed rank test uses a p-value to measure the null hypothesis that two datasets are not independent. Rejecting the null hypothesis implies that Model 11 and Model 12 are producing different returns, which would mean that the predictive modeling process described in this paper is capable of producing unique investment strategies.

The Wilcoxon signed-rank test results for comparing Model 11 and Model 12 are displayed in Figure 16. The p-values are significant at <0.0001 for 5-year and 10-year horizons across both 3-factor and 4-factor model returns, but the p-values are not significant over a 1-year horizon. These results suggest Model 11 and Model 12 are unique longer-term investment strategies, which is consistent with the long-term orientation of value investing.

**Figure 16 – Statistical Comparison Between Model 11 and Model 12**
**Wilcoxon Signed-Rank Test**

| Rolling Monthly Alpha | | | | | |
|---|---|---|---|---|---|
| Fama-French 3-Factor Model | | | Carhart 4-Factor Model | | |
| 1-year | 5-year | 10-year | 1-year | 5-year | 10-year |
| 0.155 | 0.000 | 0.000 | 0.762 | 0.000 | 0.000 |

[33] See Appendix 8 for the monthly return data (post-fees) for Models 11 and 12.

## 9. Predictive Performance of Financial Backtests

The financial backtest was performed by having each model make stock predictions on out-of-sample test data. During the predictive model development phase (prior to the backtest), predictions on out-of-sample (cross-validated) training data was used to assess predictive performance. In order to better understand the financial performance of Models 11 and 12, the predictive performance of the backtest predictions is compared with the cross-validated predictions. Performance will be compared using predictive metrics AUC (Figure 17), precision metric (Figure 18), and false negative rate metric (Figure 19). Results are shown for 1990 – 2006. Results begin with 1990 for sake of consistency with the backtest. Results after 2006 are not available because the actual outcome of Model 12's 10-year response variable is not yet available as of 12/31/16. Comparing predictive performance between Models 11 and 12 is also not uniform because both models are predicting different response variables (5-year and 10-year S&P 500 alpha). However, assessing predictive performance provides insight on potential opportunities to improve predictive performance, which could ultimately translate into improved financial performance.

Figures 17 – 19 show predictive performance is worse and more variable on the test observations compared to the cross-validated training observations. This outcome is expected. There is no lag between observations used for model training and out-of-sample cross-validated predictions, however test predictions have an *n*-period lag between observations used for model training and observations used for backtest predictions due to the use of *n*-period response variables.[34] Despite the test observation lag, Models 11 and 12 have marginally acceptable AUC and precision metrics. The range of values show improvement is needed for consistently acceptable predictive performance. Figure 19 shows both models have a high negative response rate, which means both models miss detecting many stocks that produce alpha over the desired horizon.

Figures 17 – 19 shows several performance themes. Predictive performance is worse with higher variability on the test data relative to the training data, which is a typical outcome for predictive modeling applications. Model 12 median values for AUC and precision are higher than Model 11, which suggests it is relatively easier to predict alpha over a 10-year horizon (with 10-year lagged data) than predicting alpha over a 5-yar horizon using data with a 5-year lag. In addition, the relative higher median false-negative rate for Model 12 suggests that the model is selecting fewer stocks than Model 11 at each period (and thereby missing more stocks that eventually generate alpha). The overall theme of Figures 17 – 19 is that excellent predictive performance is not necessary for the predictive models in this paper to produce statistically significant, sustained above average, financial performance; however, predictive performance has room for improvement, which could lead to even better financial performance.

---

[34] See Appendix 2 for conceptual diagram.

**Figure 17 – AUC Metric**
**Cross-Validated Training Data Vs. Test Data**

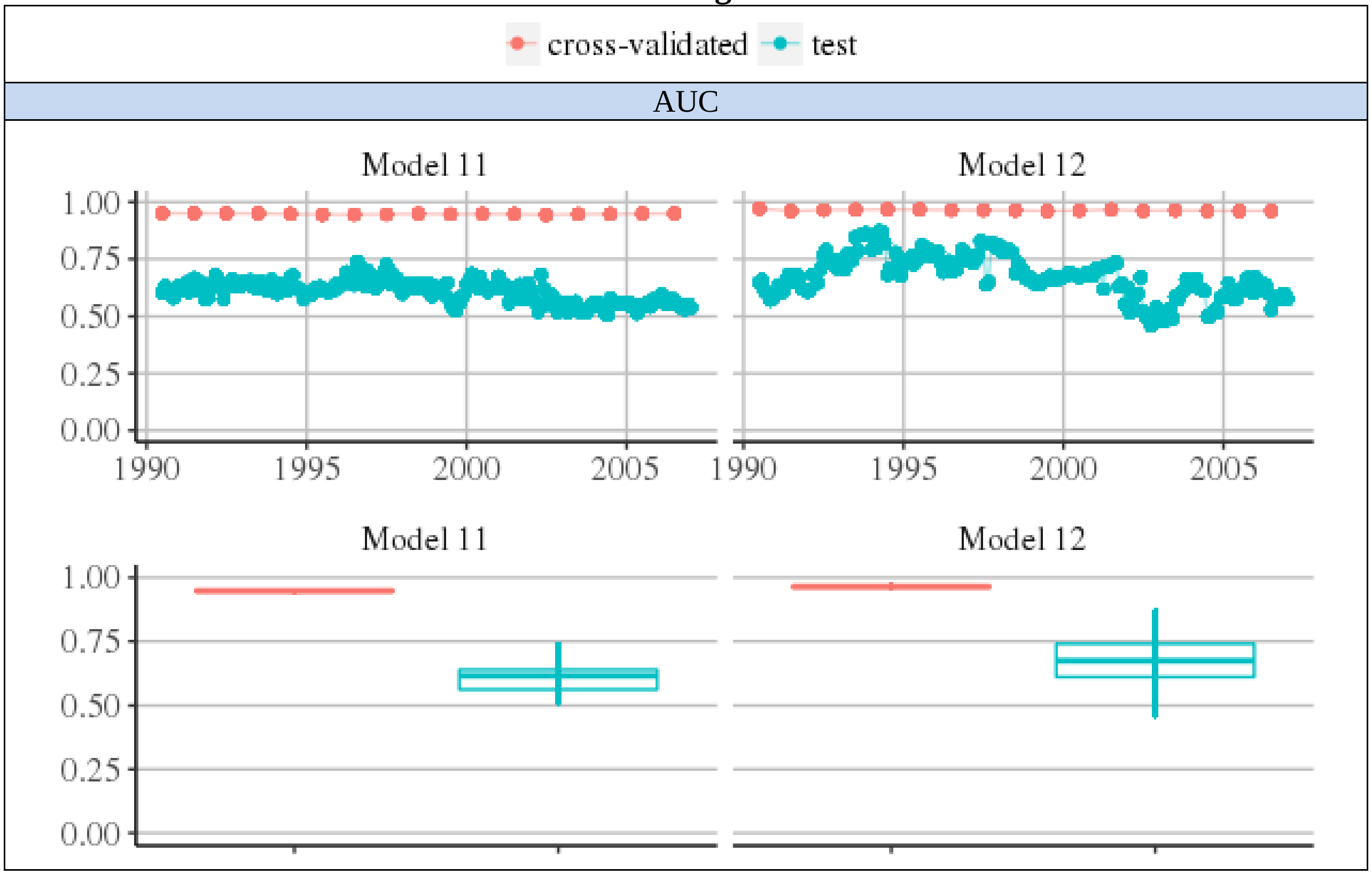

**Figure 18 – Precision Metric**
**Cross-Validated Training Data Vs. Test Data**

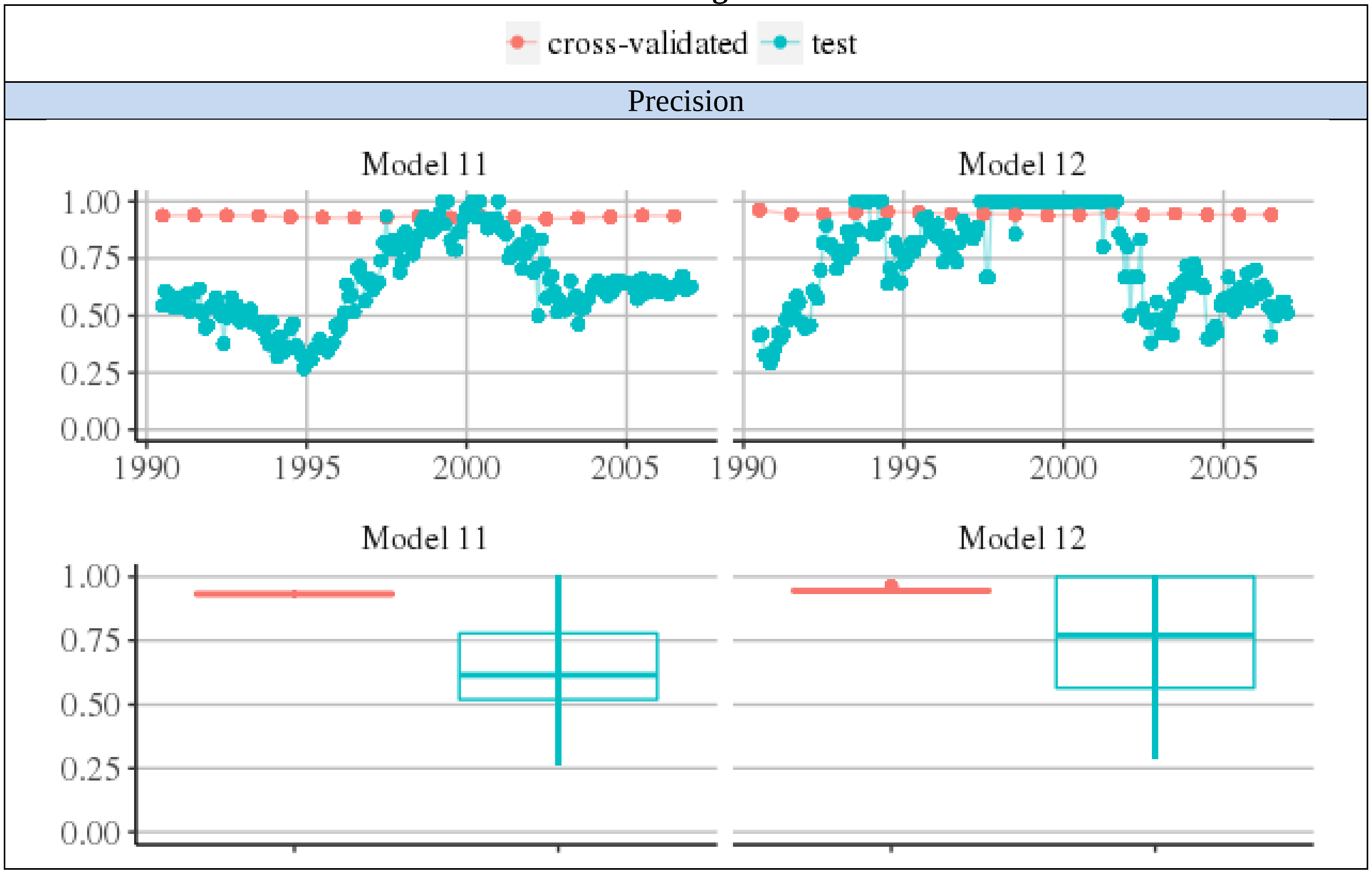

**Figure 19 – False Negative Rate Metric**
**Cross-Validated Training Data Vs. Test Data**

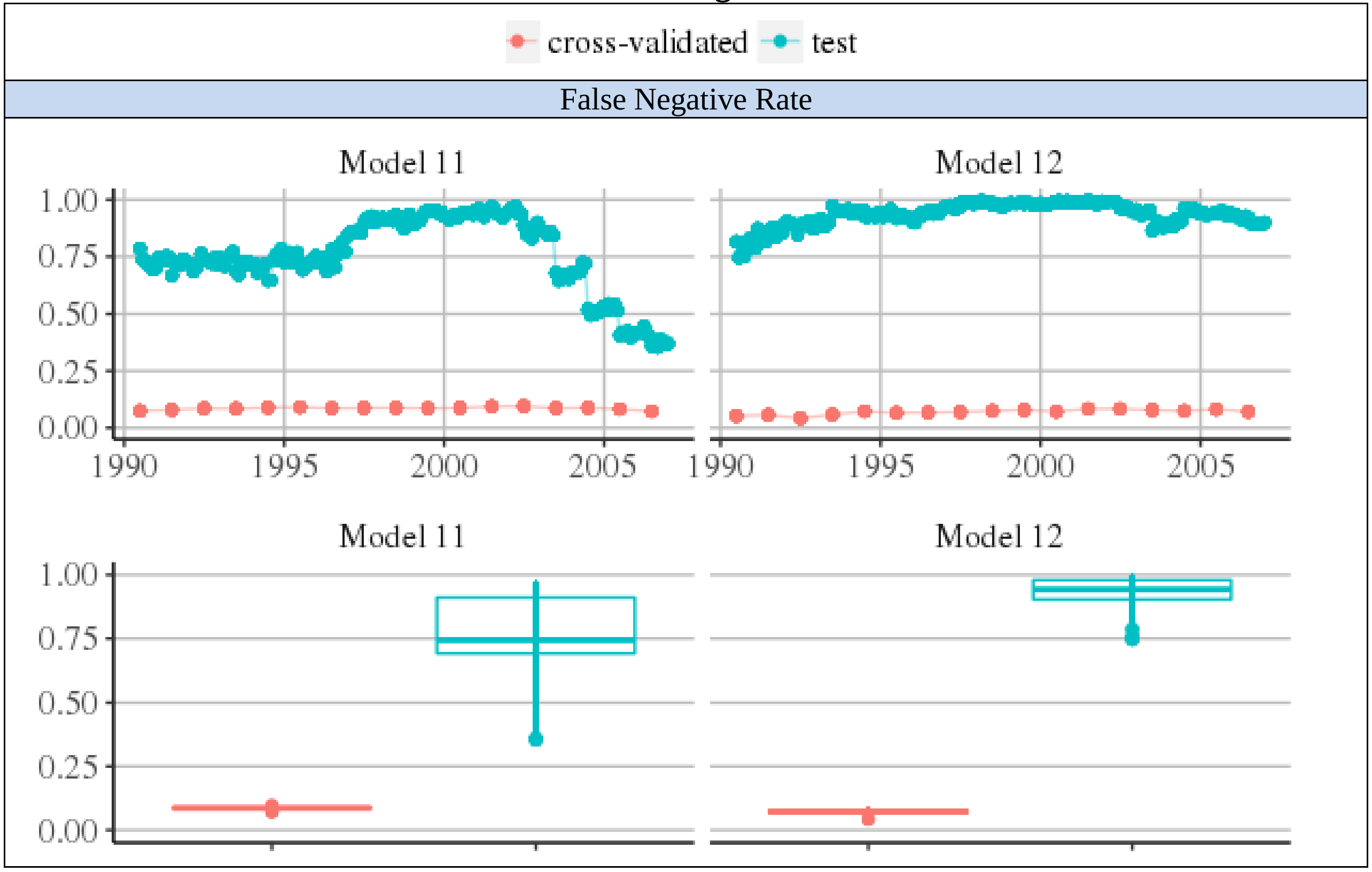

## 10. Comparing Backtest Performance Between Systematic Value Investing Strategies

The introduction of this paper claims that the dynamic and optimized nature of predictive modeling must produce results either equal or superior to non-dynamic and non-optimized (traditional) systematic value investing strategies. This claim is tested by comparing backtested results between a predictive modeling strategy and a traditional systematic value investing strategy. The comparison uses Gray & Carlisle's Quantitative Value Strategy (QV) as the systematic value investing strategy. Model 11 serves as the baseline for the predictive modeling investment strategy because it produced the best predictive and financial backtest results in the previous section. In order to ensure a valid comparison with QV, the predictor variable set for Model 11 must be changed. Model 11 used the QV Full and Buffettology variable sets as predictor variables, but QV used only the QV Full variable set. Model 11 will be retrained and backtested using only the QV Full[35] variable set as predictor variables. The updated model is named "Modified Model 11" to avoid confusion with Model 11.

Figure 20 and 21 compares backtest results for 1985 – 2016[36] between Modified Model 11 and an implementation of Gray & Carlisle's QV strategy. Most financial measures indicate that Modified Model 11 performed better than QV on most financial measures. Modified Model 11 produced statistically significant monthly returns in excess of Fama-French 3-factor model and Carhart 4-factor model monthly returns. The QV excess monthly returns were not statistically significant, however QV did producer slightly higher CAGR than Modified Model 11. Modified Model 11 also produced better Information and Sortino ratios than QV. The results suggest predictive modeling's dynamic and optimized features improve systematic value investing strategies.

[35] The QV Full variable set is used instead of the QV Basic set to give the random forest algorithm the opportunity to identify optimal relationships between the underlying variables used by Gray & Carlisle.
[36] The backtest begins in 1985 instead of 1990 because Modified Model 11 uses a response variable with a five-year lag (instead of the ten-year lag required for consistent comparisons in Figures 11 – 12).

Model 11 is also included in Figures 20 and 21 to compare the relative performance improvement between identically specified models using different sets of predictor variables. Model 11 performed better than Modified Model 11 across nearly every metric, which suggests that different predictor variable sets influence performance. Model 11 also performed better than QV, however the results are not valid evidence for suggesting the superiority of dynamic and optimized strategies because Model 11 and QV used different variable sets.

Comparing results for QV, Modified Model 11, and Model 11 is a multiple comparisons problem. The problem is addressed by applying the Bonferroni Correction to the 0.05 p-value threshold for the factor model excess returns. The results are not shown, but the monthly returns in excess of Fama-French 3-factor and Carhart 4-factor returns for Model 11 and Modified Model 11 are found to be statistically significant. The Benjamini-Hochberg procedure is not used because the Bonferroni Correction is a more conservative and definitive test.

*Quantitative Value* was published in December 2012, which publicized the QV strategy. Post-2012 comparisons between QV and predictive modeling may have bias because QV's advantages could have been arbitraged away; for example, Figure shows a post-2012 decrease in QV's rolling 5-year and 10-year factor model returns. Backtest results were also calculated for 1985 – 2012 and the comparative results between QV, Modified Model 11, and Model 11 were similar to the results previously discussed.

**Figure 20 – Backtest Results Comparison Between Systematic Value Investing and Predictive Modeling Strategies June 1985 – December 2016**

| | | Quantitative Value | Modified Model 11 | Model 11 |
|---|---|---|---|---|
| Response variable | | -- | S&P 500 5-year alpha | S&P 500 5-year alpha |
| CAGR | | 14.74% | 14.46% | 16.22% |
| Standard deviation | | 25.45% | 17.44% | 18.13% |
| Downside deviation | | 5.50% | 4.03% | 4.12% |
| Information ratio | | 0.378 | 0.532 | 0.698 |
| Sortino ratio | | 0.211 | 0.234 | 0.267 |
| Profitable months | | 56.99% | 59.89% | 60.95% |
| Best return | | 32.28% | 18.47% | 21.62% |
| Worst return | | -34.74% | -25.61% | -26.82% |
| Worst drawdown | | -66.09% | -48.20% | -47.96% |
| Cumulative drawdown | | -515,236% | -352,964% | -353,271% |
| FF 3-factor model | Alpha (annualized) | 3.32% | 3.96% | 5.19% |
| | Alpha (p-value) | 0.248 | 0.006 | 0.001 |
| | Alpha (adjusted $R^2$) | 0.613 | 0.795 | 0.788 |
| | Alpha (f-statistic) | 0.000 | 0.000 | 0.000 |
| | 1-year alpha windows | 57.34% | 70.38% | 70.65% |
| | 5-year alpha windows | 74.38% | 92.19% | 92.81% |
| | 10-year alpha windows | 90.00% | 100.00% | 100.00% |
| Carhart 4-factor model | Alpha (annualized) | 5.17% | 4.52% | 5.88% |
| | Alpha (p-value) | 0.072 | 0.002 | 0.000 |
| | Alpha (adjusted $R^2$) | 0.626 | 0.797 | 0.791 |
| | Alpha (f-statistic) | 0.000 | 0.000 | 0.000 |
| | 1-year alpha windows | 56.79% | 70.11% | 70.38% |
| | 5-year alpha windows | 74.38% | 92.18 % | 94.06% |
| | 10-year alpha windows | 91.15% | 100.00% | 100.00% |

**Figure 21 – Rolling Alpha Comparison**
**June 1985 – December 2016**

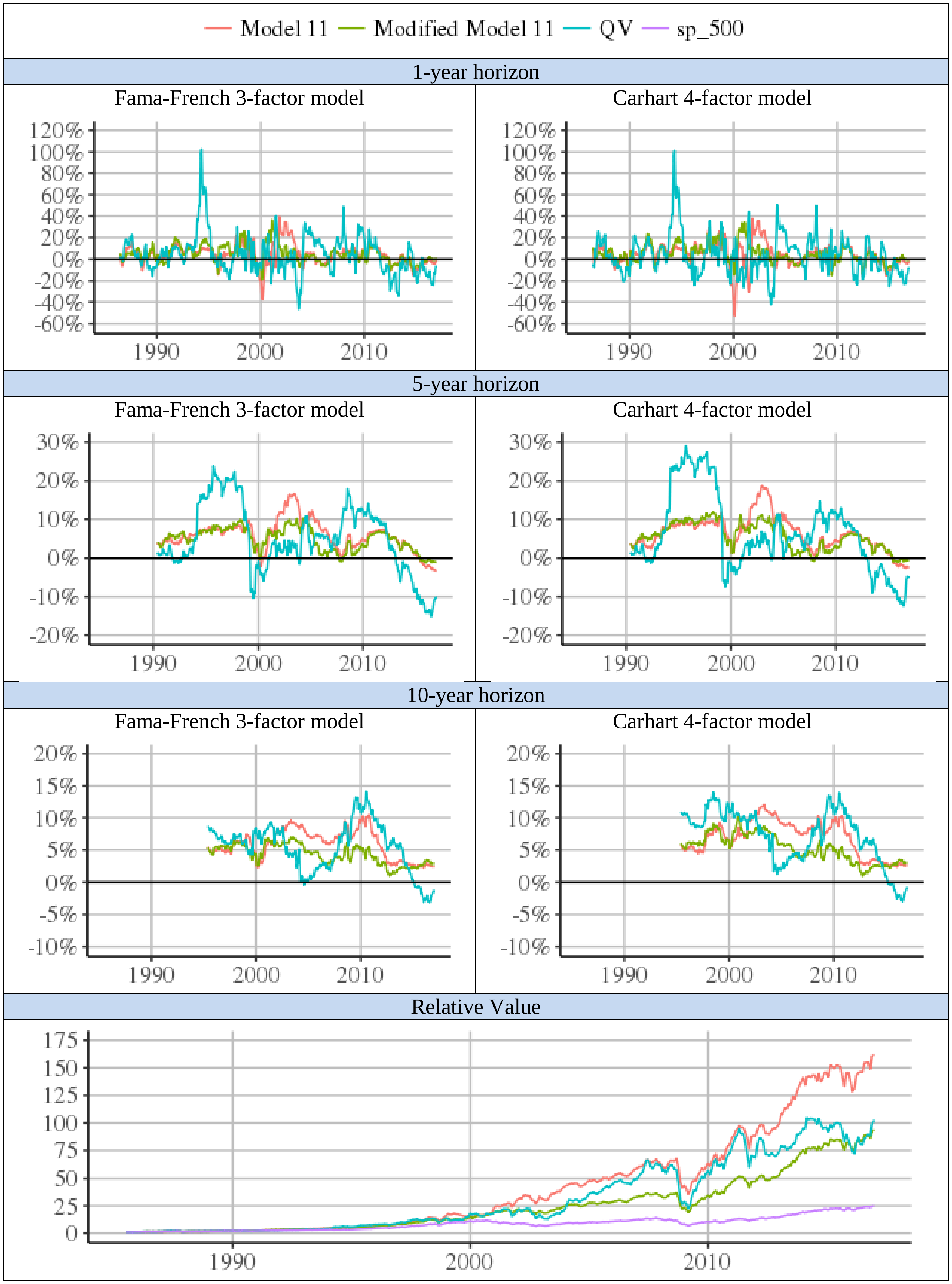

## 11. Results Discussion

Predictive modeling is occasionally discounted as data mining, where results cannot be replicated in a real-world setting. This paper uses several steps to prevent data mining. Predictive models were developed and assessed on out-of-sample cross-validated training observations. Backtested financial performance was assessed on out-of-sample test observations. The Benjamini-Hochberg procedure was used to eliminate the multiple comparisons problem for identifying statistically significant backtest portfolio returns. The Wilcoxon signed-rank test assessed whether different models represented statistically unique investment strategies. In addition, the design of the predictive modeling process prevented look-ahead bias by lagging the availability of financial statement data by six months and the dataset did not suffer from survivorship bias.

Backtest performance was compared between QV, a traditional systematic value investing strategy with non-optimized and non-dynamic features, and Modified Model 11, a predictive modeling systematic value investing strategy with optimized and dynamic features. Modified Model 11 provided arguably better performance than QV, which suggests the optimized dynamic features of predictive modeling are an improvement for traditional systematic value investing strategies. Additional backtests were performed using predictive models with different sets of predictor variables. Results showed that Model 11 and Model 12 produced statistically significant Fama-French 3-factor positive alpha and Carhart 4-factor positive alpha over a 26-year backtest. The results were definitively superior to QV and Modified Model 11, which suggests that QV does not represent a performance ceiling for systematic value investing strategies. Additional work can be performed to differentiate the effects of the expanded variable sets from the effects of using predictive models.

Model 11 and Model 12 were subject to the Benjamini-Hochberg procedure and the Wilcoxon signed-rank test. The results suggest that both models delivered statistically significant monthly returns in excess of Fama-French 3-factor model returns and Carhart 4-factor model returns over a 26-year period. The results also suggest that Model 11 and Model 12 represent different investment strategies, which infers that predictive modeling is capable of producing more than one satisfactory strategy. The results also suggest that the monthly alpha returns for Model 11 and Model 12 are not statistically different over a one-year horizon. This finding aligns with value investing’s long-term orientation.

Accepting these results should proceed with caution due to the statistical gymnastics involved. Interpreting the results as significant relies on using the Sak ratio of the expected false positive counts to infer specific models identified as significant by the Benjamini-Hochberg procedure with an expected 0.095 false positive rate. The use of the significance threshold statistic is debatable. The Bonferroni Correction is a more definitive (but excessively conservative) correction for the multiple comparison problem. The Bonferroni Correction was also performed and only Model 11 delivered excessive returns in excess of both factor models over the 26-year period. Despite the debatable interpretation of the portfolio backtest results, the predictive modeling techniques presented in this paper were chosen for their relative simplicity and ease of conceptual understanding. Using techniques that are more sophisticated will likely improve predictive and financial performance. In summary, the results presented in this paper are not definitive, but they are strong enough to warrant further exploration of using predictive modeling as a systemic value investing strategy.

The backtest revealed that models trained on "S&P 500 *n*-year alpha" response variables[37] performed better overall than models trained on "Fama-French *n*-year alpha" response variables[38]. This phenomenon was not formally investigated. However, one explanation is the S&P 500 models performed better because value, size, and market risk factors are not regressed out from the response variables. The predictor variable sets included several metrics that used or measured each stock's market capitalization, value, and stock price volatility. Perhaps the inclusion of factor information in the response variable benefitted the predictive ability of the financial metrics, which resulted in stock picks that were not biased towards market risk, value, or size factors. The Fama-French models were unable to control for factor exposure as effectively because the factor signals had been removed from the response variables.

Finally, the predictive performance of backtest predictions contrasted with the financial performance. Model 12 had relatively better predictive performance than Model 11, but the result was relatively worse financial performance. This was also not formally investigated. One possibility is that although Model 12 had relatively better binary predictions, perhaps the Model 12 incorrect predictions had materially worse financial performance than incorrect predictions for Model 11.

## 12. Extensions

The predictive modeling process in this paper is a conceptual introduction to a predictive modeling methodology. Simplicity was prioritized over performance wherever possible to avoid unnecessary technical discussion for introductory purposes. Methods that are more complex may produce performance improvements with potentially definitive statistical significance. Other methods include exploring different predictor variable sets, predictive algorithms, variable engineering techniques, and missing data imputation techniques. Exploratory work can help understand the predictive/financial performance trade-off demonstrated by Model 11 and Model 12. Perhaps using quantitative or compound (binary above a certain threshold) response variables would provide insights. In addition, instead of using an iterative approach for developing models, all possible model permutations could be considered. Data mining can be controlled using the procedures outlined in this paper.

---

[37] Model 6, Model 8, Model 9
[38] Model 10, Model 11, Model 12

Additional financial work might explore the financial characteristics of portfolios produced by predictive models including examining portfolio performance using different thresholds for market liquidity. Other topics include assessing whether predictive models discover new return factors and exploring factor persistence when using a systematic investing strategy with dynamic features.

Using predictive modeling as an active investment strategy is feasible. The models and code in this paper were written in the statistical programming language R. The models outputs lists of securities to buy (not shown). The lists were used for building backtested portfolios, and they can just as easily be used for real-time stock purchasing. Similarly, the models in this paper can be reproduced by applying the specifications for a desired model (Appendix 3) to the data preparation discussed in Section 4. Exhaustive implementation notes are also noted in Appendices 5 – 8. Reproducing the models requires purchasing financial data from Compustat and stock price data from the Center for Research in Security Prices (CRSP). The computational requirements and time costs for the modeling process described in this paper are not excessive and do not depend on a high performance computing environment.

## 13. Conclusion

This paper contributes to the literature on systematic value investing by defining value investing as an optimization problem and introducing predictive modeling as a general method for adding dynamic optimization features to a systematic value investing strategy. Results from a 31-year backtest comparison showed that predictive modeling's dynamic and optimization features produced better results than a non-optimized, non-dynamic systematic investing strategy using the same set of financial metrics. A 26-year backtest using a larger set of financial metrics suggest that a predictive modeling process generated two statistically unique investment strategies with statistically significant Fama-French 3-factor model and Carhart 4-factor model alphas. All results were financially satisfactory and produced using relatively simple modeling methods, which suggest additional improvements are possible. The results may motivate practitioners to further explore predictive modeling as a systematic value investing strategy. Researchers may find benefit in using predictive modeling as tool to continue exploring the behavior and nature of stock market returns.

**Appendix 1 – QV Methodology**

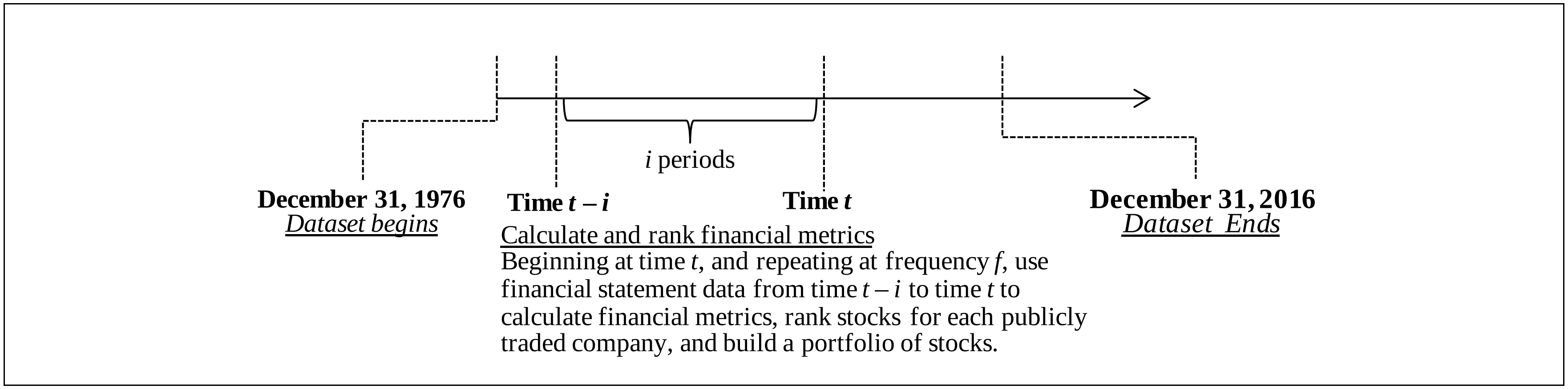

**Appendix 2 – Predictive Modeling Methodology**

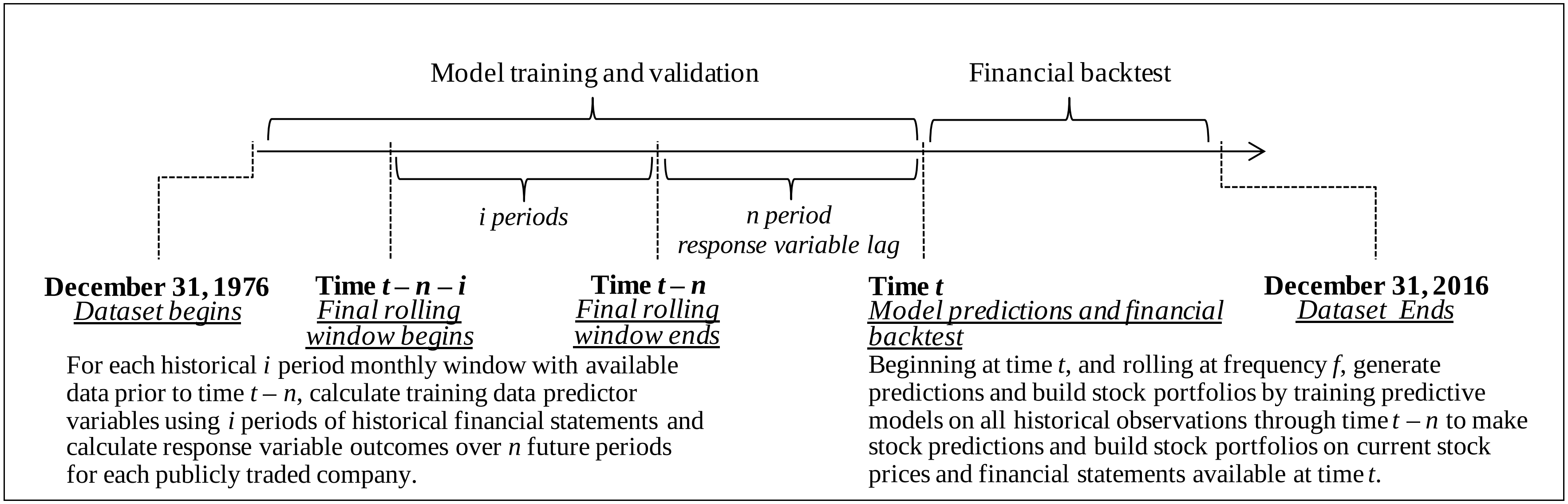

**Appendix 3 – Predictive Model Specifications**

| Model | Algorithm | Predictor variable set | Stock selection | Response variable |
|---|---|---|---|---|
| 1 | Logistic Regression | QV Basic | > 50% | FF 1-year alpha |
| 2 | Classification Tree | | | |
| 3 | Random Forest | | | |
| 4 | | QV Full | | |
| 5 | | Buffettology | | |
| 6 | | Buffettology & QV Full | | |
| 7 | | | Top decile | |
| 8 | | | > 50% | FF 5-year alpha |
| 9 | | | | FF 10-year alpha |
| 10 | | | | 1-year alpha |
| 11 | | | | 5-year alpha |
| 12 | | | | 10-year alpha |

**Appendix 4 – Overview of Predictive Algorithms**

This is a conceptual introduction to each predictive algorithm. Full expositions for each algorithm are outside the scope of this paper. See Friedman, Hastie, & Tibshirani (2009) and James, Witten, Hastie, & Tibshirani for technical treatment and bibliographies.

Logistic regression is a sub-class of generalized linear models (GLMs). GLMs are algebraic equations that use coefficients to describe the relationship between a set of independent (predictor) variables and a dependent (response) variable. GLMs can be specified for different purposes, including predicting quantities (linear regression) or categories (logistic regression categories). Specific GLMs produce output predictions in the same scale as the dependent variable (continuous variable, binary variable, etc.) by using a "link function" that accounts for the cumulative distribution of response variable values. GLM models estimate relationship between a set of independent variables and a dependent variable by searching for the set of independent variable coefficients that minimize the distance between the actual (known) response variable values and the values predicted by the GLM model. The resulting set of independent variable coefficients is then used for making predictions where independent variable data is available, but the response variable is unknown.

Classification trees apply a series of binary decisions (called "decision nodes") about the values of independent variables. All classification tree paths eventually terminate with an "end node" that assigns a predefined outcome (or value) to the response variable. Classification trees can be visualized as hierarchical structures that resemble an upside-down "tree" with many interconnected decision nodes (branches) that result in many end-nodes (leafs). Classification trees are "grown" by setting default values for the number of independent variables, decision nodes, and end nodes to include in the "tree building" process. A classification tree algorithm iteratively assesses different combinations of independent variables, decision notes, and end node values to identify the classification tree specification that produces response variable values that are nearest to the actual values. The resultant classification tree is then used for making predictions where independent variable data is available, but the response variable is unknown.

Random forests are conceptually similar to classification trees. Instead of using a single classification tree to output a prediction, random forests grow many distinct classification trees (a "forest" of trees). Random forest algorithms predict a response variable value by aggregating all the predictions from the individual decisions trees using either the median/mean value (for quantitative predictions) or the mode value (for categorical predictions). Random forests are "random" because each classification tree is grown using a different random sample of training observations and each classification tree may also use different random subsets of independent variables.

## Appendix 5 – Quantitative Value Variables Overview

The following variables are included in the "QV Full" predictor variable set. Predictor variables used for the "QV Basic" variable set are noted.

| Quantitative Value Checklist | | Variable | Description | Variable Set | Source |
|---|---|---|---|---|---|
| **1** | **Avoid stocks at risk of sustaining a permanent loss of capital** | | | | |
| 1.1 | Identify potential frauds and manipulators | | | | |
| 1.1.1 | Accrual screens | sta | Scaled total assets | QV Full | Sloan |
| | | sta_pct | Scaled total assets (cumulative percentile) | QV Full | |
| | | snoa | Scaled net operating assets | QV Full | Gray & Carlisle |
| | | snoa_pct | Scaled net operating assets (cumulative percentile) | QV Full | |
| | | comboaccrual | Average of sta_pct, snoa_pct | QV Basic | Gray & Carlisle |
| 1.1.2 | Fraud and manipulation screens | dsri | Days' sales in receivables index | QV Full | Beneish |
| | | gmi | Gross margin index | QV Full | |
| | | aqi | Asset quality index | QV Full | |
| | | sgi | Sales growth index | QV Full | |
| | | depi | Depreciation index | QV Full | |
| | | sgai | Sales, general, and administrative expenses index | QV Full | |
| | | lvgi | Leverage index | QV Full | |
| | | tata | Total accruals to total assets | QV Full | |
| | | pman | Probability of manipulation | QV Basic | |
| 1.2 | Identify stocks at high risk of financial distress | | | | |
| 1.2.1 | Probability of financial distress | nimtaavg | Net income to market-valued total assets (lagged values) | QV Full | Campbell, Hilscher, & Szilagyi |
| | | tlmta | Total liabilities to market-value total assets | QV Full | |
| | | cashmta | Cash and short-term assets to market-value total assets | QV Full | |
| | | exretavg | Monthly log excess return relative to S&P 500 | QV Full | |
| | | return_3mos_sd | Standard deviation of daily stock return over the past 3 months | QV Full | |
| | | rsize | Log ratio of market capitalization to S&P 500 | QV Full | |
| | | mb | Market to book ratio | QV Full | |
| | | price_log | Log adjusted close share price truncated at $15 | QV Full | |
| | | pfd | Probability of financial distress (12 month lag) | QV Basic | |
| 1.3 | Eliminate stocks at risk of sustaining a permanent loss of capital | | | | |
| Implementation step. No variables added | | | | | |
| **2** | **Find the cheapest stocks** | | | | |
| 2 | Find the cheapest stocks | price_cheap | Earnings yield | QV Basic | Greenblat |
| **3** | **Find highest-quality stocks** | | | | |
| 3.1 | Franchise power | roa | Return on assets | QV Full | Gray & Carlisle |
| | | roa_8_gmean | Eight year return on assets | QV Full | |
| | | roa_8_gmean_pct | Eight year return on assets (cumulative percentile) | QV Full | |
| | | roc_8_gmean | Eight year return on capital | QV Full | |
| | | fcfa | Long-term free cash flow on assets | QV Full | |
| | | fcfa_pct | Long-term free cash flow on assets (cumulative percentile) | QV Full | |
| | | gm_8_gmean | Eight year margin growth | QV Full | |
| | | gm_8_gmean_pct | Eight year margin growth (cumulative percentile) | QV Full | |
| | | gms | Eight year gross margin stability | QV Full | |
| | | gms_pct | Eight year gross margin stability (cumulative percentile) | QV Full | |
| | | margin_max | See Gray & Carlisle for description | QV Full | |
| | | fran_power | See Gray & Carlisle for description | QV Full | |
| 3.2 | Financial strength | | | | |
| 3.2.1 | Current profitability | roa | Return on assets | QV Full | Piotroski and Gray & Carlisle |
| | | fs_roa | Indicator variable for roa > 0 | QV Full | |
| | | fcfta | Free cash flow to total assets | QV Full | |
| | | fs_fcfta | Indicator variable for fcfta > 0 | QV Full | |
| | | accrual | fcfta – roa | QV Full | |
| | | fs_accrual | Indicator variable for accrual > 0 | QV Full | |
| 3.2.2 | Stability | lever | See Piotroski and Boudoukh, Michaely, Richardson, & Roberts for | QV Full | Piotroski and Boudoukh, |
| | | | description | | Michaely, Richardson, & Roberts |
| | | fs_lever | Indicator variable for lever > 0 | QV Full | |
| | | liquid | See Piotroski and Boudoukh, Michaely, Richardson, & Roberts for description | QV Full | |
| | | fs_liquid | Indicator variable for liquid > 0 | QV Full | |
| | | neqiss | See Piotroski and Boudoukh, Michaely, Richardson, & Roberts for description | QV Full | |
| | | fs_neqiss | Indicator variable for neqiss > 0 | QV Full | |
| 3.2.3 | Recent operational improvements | roa_change | Year-over-year change in return on assets | QV Full | Gray & Carlisle |
| | | fs_roa | Indicator variable for roa_change > 0 | QV Full | |
| | | fcfta_change | Year-over-year change in free cash flow to total assets | QV Full | |
| | | fs_fcfta_change | Indicator variable for fcfta_change > 0 | QV Full | |
| | | gm_change | Year-over-year change in gross margin | QV Full | |
| | | fs_gm | Indicator variable for gm_change > 0 | QV Full | |
| | | turnover_change | Year-over-year change in asset turnover | QV Full | |
| | | fs_turnover | Indicator variable for turnover_change > 0 | QV Full | |
| 3.2.4 | Financial strength | fin_strength | See Gray & Carlisle for description | QV Full | Gray & Carlisle |
| 3.3 | Identify quality | quality | See Gray & Carlisle for description | QV Basic | Gray & Carlisle |

Implementation note: The following variables were removed from the "QV Full" predictor variable set due to collinearity (high correlation) with other variables in the QV Full set. When predictor variables are highly correlated, the logistic regression predictive algorithm is unable to estimate variable coefficients because the covariance matrix has infinite solutions. Although the collinearity issue does not affect the classification tree or random forest algorithms, the following variables were removed from the "QV Full" predictor variable set for all models to ensure consistent comparisons between predictive algorithms (Models 1 – 3) as well as for comparison with subsequent models.

| Removed Variables | |
|---|---|
| Variable | Reason for removal |
| fin_strength | High correlation w/ quality |
| fran_power | High correlation with fran_power_pct |
| gm_8_gmean_pct | High correlation with gm_8_gmean |
| gms | High correlation w/ gms_pct |
| margin_max | High correlation w/ gm_8_gmean_pct |
| roa_8_gmean_pct | High correlation with roa_8_gmean |
| snoa_pct | High correlation with snoa |
| sta | Linear combination with tata |
| sta_pct | Linear combination with tata |
| turnover_change | High correlation with fcfa_pct |

## Appendix 6 – Quantitative Value Variable Details

Note: Unless otherwise noted, variable names refer to variables names used on the Wharton Research Data Services (WRDS) web query form for the CCM database.

### Checklist Item 1.1.1 – Accrual Screens

**Formulas**

$$sta_t = \frac{ca_t - cl_t - dp_t}{at_t}$$

$$sta_pct_t = Percentile_{cum}(sta)$$

$$snoa_t = \frac{oa_t - ol_t}{at_t}$$

$$snoa_pct_t = Percentile_{cum}(snoa)$$

$$comboaccrual_t = \frac{snoa_pct_t + sta_pct_t}{2}$$

$where$:

$$ca_t = (act_t - act_{t-1}) - (che_t - che_{t-1})$$

$$cl_t = (lct_t - lct_{t-1}) - (dd1_t - dd1_{t-1}) - (txp_t - txp_{t-1})$$

$$oa_t = (at_t - che_t)$$

$$ol_t = (at_t - dle_t - dltt_t - mib_t - pstk_t - ceq_t)$$

**Implementation Notes**

For variables ca, cl, sta, oa, ol, snoa, at:

- Treat infinite values as a missing value.
- Remove observations with 100% missing values.
- Impute missing values with last non-missing observation.

### Checklist Item 1.1.2 – Fraud and manipulation screens

**Formulas**

$$pman_t = P(probm_t)$$

$where$:

$$probm_t = -4.84 + (0.92 * dsri_t) + (0.528 * gmi_t) + (0.404 * aqi_t) + (0.892 * sgi_t) + (0.115 * depi_t) - (0.172 * sgai_t) + (4.679 * tata_t) - (0.327 * lvgi_t)$$

$$dsri_t = \frac{\left(\frac{rect_t}{sale_t}\right)}{\left(\frac{rect_{t-1}}{sale_{t-1}}\right)}$$

$$gmi_t = \frac{\left(\frac{sale_t - cogs_t}{sale_t}\right)}{\left(\frac{sale_{t-1} - cogs_{t-1}}{sale_{t-1}}\right)}$$

$$aqi_t = \frac{\left(1 - \left(\frac{act_t + ppent_t}{at_t}\right)\right)}{\left(1 - \left(\frac{act_{t-1} + ppent_{t-1}}{at_{t-1}}\right)\right)}$$

$$sgi_t = \frac{sale_t}{sale_{t-1}}$$

$$depi_t = \frac{\left(\frac{dp_{t-1} - am_{t-1}}{dp_{t-1} - am_{t-1} + ppent_{t-1}}\right)}{\left(\frac{dp_t - am_t}{dp_t - am_t + ppent_t}\right)}$$

$$sgai_t = \frac{\left(\frac{xsga_t}{sale_t}\right)}{\left(\frac{xsga_{t-1}}{sale_{t-1}}\right)}$$

$$lvgi_t = \frac{\left(\frac{dltt_t + lct_t}{at_t}\right)}{\left(\frac{dltt_{t-1} + lct_{t-1}}{at_{t-1}}\right)}$$

$$tata_t = \frac{\left((act_t - act_{t-1}) - (che_t - che_{t-1}) - (lct_t - lct_{t-1}) - (dd1_t - dd1_{t-1}) - (txp_t - txp_{t-1}) - dp_t\right)}{at_t}$$

**Implementation Notes – dsri**

- When $rect_{t-1} = 0$, set $dsri_t = rect_t/sale_t$ to avoid zero divisor error.
- When $sale_t = 0$, set $dsri_t = rect_t$ to avoid zero divisor error.
- When $sale_t = 0$ and $sale_{t-1} = 0$, set $dsri_t = rect_t/rect_{t-1}$
- When $rect_t = 0$ and $rect_{t-1} = 0$, set $dsri_t = 0$
- When $rect_t = 0$ and $sale_t = 0$, set $dsri_t = 0$
- When $sale_t = 0$ and $rect_{t-1} = 0$, set $dsri_t = rect$
- When $sale_{t-1} = 0$ and $rect_{t-1} = 0$, set $dsri_t = rect_t/sale_t$
- When $sale_t = 0$ and $sale_{t-1} = 0$ and $rect_{t-1} = 0$, set $dsri_t = rect_t$
- When $sale_t = 0$ and $sale_{t-1} = 0$ and $rect_t = 0$ and $rect_{t-1} = 0$, set $dsri_t = 1$
- Treat infinite values as a missing value.
- Remove observations with 100% missing values.
- Impute missing values with last non-missing observation.

**Implementation Notes – gmi**

- When $sale_t = 0$ and $sale_{t-1} = 0$ and $cogs_t = 0$ and $cogs_{t-1} = 0$, set $gm_t = 1$
- When $sale_t = 0$ and $sale_{t-1} = 0$ and $cogs_t \neq 0$ and $cogs_{t-1} \neq 0$, set $gm_t = 1$

- When $sale_t = 0$ and $sale_{t-1} = 0$ and $cogs_t \neq 0$ and $cogs_{t-1} = 0$, set $gm_t = 1$
- When $sale_t = 0$ and $sale_{t-1} = 0$ and $cogs_t = 0$ and $cogs_{t-1} \neq 0$, set $gm_t = 1$
- When $sale_t = 0$ and $sale_{t-1} \neq 0$ and $cogs_t = 0$ and $cogs_{t-1} \neq 0$, set $gm_t = 0$
- When $sale_t = 0$ and $sale_{t-1} \neq 0$ and $cogs_t = 0$ and $cogs_{t-1} = 0$, set $gm_t = 0$
- When $sale_t \neq 0$ and $sale_{t-1} = 0$ and $cogs_t = 0$ and $cogs_{t-1} \neq 0$, set $gm_t = sale_t + 1$
- When $sale_t = 0$ and $sale_{t-1} = 0$ and $cogs_t \neq 0$ and $cogs_{t-1} = 0$, set $((gm_t = sale_t – cogs_t)/sale_t) + 1$
- Treat infinite values as a missing value.
- Remove observations with 100% missing values.
- Impute missing values with last non-missing observation.

**Implementation Notes – aqi**

- When $aqi_t$ = missing and $1 - ((act_{t-1} + ppent_{t-1}) / at_{t-1}) = 0$ and $1- ((act_t + ppent_t) / at_t) = 0$, set $aqi_t = 1$
- When $aqi_t$ = missing and $((act_{t-1} + ppent_{t-1}) / at_{t-1}) = 0$, set $aqi_t = 1- ((act_t + ppent_t) / at_t)$
- When $act_{t-1} = 0$ and $ppent_{t-1} = 0$ and $at_{t-1} = 0$, set $aqi_t = 1- ((act_t + ppent_t) / at_t)$
- When $act_t = 0$ and $ppent_t = 0$ and $at_t = 0$, set $aqi_t = 1$
- When $act_t = 0$ and $ppent_t = 0$ and $at_{t-1} = 0$ and $act_{t-1} = 0$ and $ppent_{t-1} = 0$ and $at_{t-1} =0$, set $aqi_t = 1$
- Treat infinite values as a missing value.
- Remove observations with 100% missing values.
- Impute missing values with last non-missing observation.

**Implementation Notes – sgi**

- When $sale_t = 0$ and $sale_{t-1} =0$, set $sgi_t =1$
- When $sale_t \neq 0$ and $sale_{t-1} =0$, set $sgi_t = sale_t + 1$
- Treat infinite values as a missing value.
- Remove observations with 100% missing values.
- Impute missing values with last non-missing observation.

**Implementation Notes – depi**

- When $dp_{t-1} = 0$ and $am_{t-1} = 0$ and $ppent_{t-1} = 0$ and $dp_t = 0$ and $am_t = 0$ and $ppent_t = 0$, set $depi_t = 1$
- When $dp_{t-1} = 0$ and $am_{t-1} = 0$ and $ppent_{t-1} \neq 0$ and $dp_t = 0$ and $am_t = 0$ and $ppent_t \neq 0$, set $depi_t = 1$
- When $dp_{t-1} = 0$ and $am_{t-1} = 0$ and $ppent_{t-1} = 0$ and $dp_t = 0$ and $am_t \neq 0$ and $ppent_t = 0$, set $depi_t = 2$
- When $dp_{t-1} \neq 0$ and $am_{t-1} \neq 0$ and $ppent_{t-1} \neq 0$ and $dp_t = 0$ and $am_t = 0$ and $ppent_t = 0$, set $depi_t = ((dp_{t-1} - am_{t-1})/ (dp_{t-1} - am_{t-1} + ppent_{t-1})) + 1$
- When $dp_{t-1} = 0$ and $am_{t-1} = 0$ and $ppent_{t-1} = 0$ and $dp_t \neq 0$ and $am_t \neq 0$ and $ppent_t \neq 0$, set $depi_t = 0$
- When $dp_{t-1} = 0$ and $am_{t-1} = 0$ and $ppent_{t-1} = 0$ and $dp_t \neq 0$ and $am_t = 0$ and $ppent_t \neq 0$, set $depi_t = 0$
- When $dp_{t-1} = 0$ and $am_{t-1} = 0$ and $ppent_{t-1} = 0$ and $dp_t = 0$ and $am_t = 0$ and $ppent_t \neq 0$, set $depi_t = 0$
- When $depi_t$ = missing and $dp_t = am_t$, set $depi_t = ((dp_{t-1} - am_{t-1})/ (dp_{t-1} - am_{t-1} + ppent_{t-1})) + 1$
- Treat infinite values as a missing value.
- Remove observations with 100% missing values.
- Impute missing values with last non-missing observation.

**Implementation Notes – sgai**

- When $sale_t < 1$, set $sale_t = 1$ for sgai calculations only.
- When $xsga_t = 0$ and $xsga_{t-1} = 0$, set $sgai_t = 0$
- When $xsga_{t-1} = 0$, set $sgai_t = xsga_t / sale_t$
- Treat infinite values as a missing value.
- Remove observations with 100% missing values.
- Impute missing values with last non-missing observation.

**Implementation Notes – lvgi**

- When $dltt_{t-1}$ = 0 and $lct_{t-1}$ = 0, set $lvgi_t$ = (($dltt_t$ + $lct_t$) / $at_t$) + 1
- When $dltt_{t-1}$ = 0 and $lct_{t-1}$ = 0 and $at_t$ = 0, set $lvgi_t$ = $dltt_t$ + $lct_t$ + 1
- Treat infinite values as a missing value.
- Remove observations with 100% missing values.
- Impute missing values with last non-missing observation.

**Implementation Notes – tata**

- Treat infinite values as a missing value.
- Remove observations with 100% missing values.
- Impute missing values with last non-missing observation.

## Checklist Item 1.2.1.1 – Probability of financial distress

**Formulas**

$$pfd_t = \frac{1}{1 + e^{-lpfd_t}}$$

$$lpfd_t = -9.16 - (20.26 * nimtaavg_t) + (1.42 * tlmta_t) - (7.13 * exretavg_t) + (1.41 * return_3mos_sd_t) - (0.045 * rsize_t) - (2.13 * cashmta_t) + (0.075 * mb_t) - (0.058 * price_log_t)$$

$where$

$$nimtaavg_t = \frac{1- \Phi^3}{1- \Phi^{12}} * (\Phi^0 * nimta_t) + (\Phi^3 * nimta_{t-3}) + (\Phi^6 * nimta_{t-6}) + (\Phi^9 * nimta_{t-9})$$

$$\Phi = \frac{1}{\sqrt[3]{2}}$$

$$nimta_t = \frac{niq_t}{adj_close_t * shrout_t + ltq_t}$$

$$tlmta_t = \frac{ltq_t}{adj_close_t * shrout_t + ltq_t}$$

$$exretavg_t = \frac{1- \Phi^1}{1- \Phi^{12}} * \sum_{0}^{-11} \left(\Phi^{abs(t)} * exret_t\right)$$

$$exret_t = ln\,(1.01 + return_monthly_t) - ln(1.01 + sp500_return_monthly_t)$$

$$return_monthly_t = \frac{(adj_close_t + div_t)}{adj_close_{t-1}}$$

$$div_t = \frac{divamt_t}{cfacpr_t}$$

$$sp500_return_monthly_t = \left( \prod_{0}^{-23} 1 + vwretd_t \right) - 1$$

$$return_3mos_sd_t = std[return_daily_t \ldots return_daily_{t-62}]$$

$$return_daily_t = \left( \frac{(adj_close_t + div_t)}{adj_close_{t-1}} \right) - 1$$

$$rsize_t = ln\left(1 + \frac{shrout_t + adj_close_t}{usdval_t}\right)$$

$$cashmta_t = \frac{cheq_t}{adj_close_t * shrout_t + ltq_t}$$

$$mb_t = \frac{mv_t}{bv_adj_t}$$

$$bv_adj_t = \big((mv_t - bv_t) * 0.1\big) + bv_t$$

$$bv_t = txdb_t + itcb_t + prca_t - pstkrv_t + seq_t$$

$$mv_t = adj_close_t * shrout_t$$

$$price_log_t = \begin{cases} ln(15) \text{ if } adj_close_t > 15 \\ ln(adj_close_t) \text{ otherwise} \end{cases}$$

**Implementation Notes – all variables**

- Campbell, Hilscher, & Szilagyi lag the availability of all variables by two months. This practice is not followed because all financial statement variables are lagged by six months for sake of consistency with Gray & Carlisle.

**Implementation Notes – return_3mos_sd**

- When $daily_return_t$ is infinite, set $daily_return_t$ = missing
- When $adj_close_t = 0$ or missing and $adj_close_{t-1} = 0$ or missing, set $daily_return_t = 0$
- When $adj_close_t \neq 0$ or missing and $adj_close_{t-1} = 0$ or missing, set $daily_return_t = adj_close_t + divamt_t$
- When $adj_close_t = 0$ or missing and $adj_close_{t-1} \neq 0$ or missing, set $daily_return_t = -1$
- Treat infinite values as a missing value
- Remove observations with 100% missing values
- Impute missing values with last non-missing observation
- Define "3 months of daily returns" as 63 prior observations (daily)
- Remove observations where > 108 calendar days elapsed between prior 63 observations
- Windsorize values between the $10^{th}$ percentile and $90^{th}$ percentile rolling cumulative values

**Implementation Notes – exretavg**

- Add cumulative dividends to share price for return calculation
- When $adj_close_{t-1} = 0$, set $monthly_return_t = adj_close_t$
- Monthly excess return is calculated as ln(1.01 + return_monthly) - ln(1.01 + vwretd_return_monthly).

Returns are transformed by +1.01 to avoid natural logarithm calculations producing NA and zero values.

- When vwretd is missing, set vwretd = 0
- As suggested by Campbell, Hilscher, & Szilagyi, when exret is missing, set exret to the cross sectional mean value of exret for the current date.
- Windsorize values between the 10th percentile and 90th percentile rolling cumulative values

**Implementation Notes – rsize**

- The log ratio of market capitalization to market size (rsize) is calculated as ln( 1 + (shrout*adj_close) / usdval). The ratio is transformed by +1 to avoid natural logarithm calculations producing NA and zero values when market capitalization = 0.
- Treat infinite values as a missing value
- Remove observations with 100% missing values
- usdval units were converted from thousands to millions
- Impute missing values with last non-missing observation
- Windsorize values between the 10th percentile and 90th percentile rolling cumulative values

**Implementation Notes – nimtaavg**

- Treat infinite values as a missing value
- Remove observations with 100% missing values
- Impute missing values with last non-missing observation
- Windsorize values between the 10th percentile and 90th percentile rolling cumulative values

**Implementation Notes – tlmta**

- Treat infinite values as a missing value
- Remove observations with 100% missing values
- Impute missing values with last non-missing observation
- Windsorize values between the 10th percentile and 90th percentile rolling cumulative values

**Implementation Notes – cashmta**

- Treat infinite values as a missing value
- Remove observations with 100% missing values
- Impute missing values with last non-missing observation
- Windsorize values between the 10th percentile and 90th percentile rolling cumulative values

**Implementation Notes – mb**

- As suggested by Campbell, Hilscher, & Szilagyi, when $book_value_t$ is unavailable, it is substituted with book value as defined by Cohen, Polk, & Vuolteenaho (2003).
- Treat infinite values as a missing value
- Remove observations with 100% missing values
- Impute missing values with last non-missing observation
- When $mv_t = 0$ and $bv_adj_t = 0$, set mb = 0
- Windsorize values between the 10th percentile and 90th percentile rolling cumulative values
- When $txdb_t$ is missing, set $txdb_t = 0$
- When $itcb_t$ is missing, set $itcb_t = 0$
- When $prca_t$ is missing, set $prca_t = 0$
- When $pstkrv_t$ is missing, set $pstkrv_t = pstkl_t$
    - When $pstkl_t$ is missing, set $pstkl_t = pstk_t$
    - When $pstk_t$ is missing, set $pstk_t = 0$

- When $seq_t$ is missing, set $seq_t = ceq_t + pstk_t$
- When $ceq_t$ or $pstk_t$ is missing, set $seq_t = at_t – lt_t$
  - When $at_t$ or $lt_t$ is missing, set $seq_t = 0$

## Checklist Item 2 – Find the cheapest stocks

**Formula**

$$tev_t = (adj_close_t * shrout_t) + (dltt_t + dlc_t) - (ch_t + at_t - lct_t) + pstkrv_t + mib_t$$

**Implementation Notes**

- When $pstkrv_t$ is missing, set $pstkrv_t = pstkl_t$
  - When $pstkl_t$ is missing, set $pstkl_t = pstk_t$
  - When $pstk_t$ is missing, set $pstk_t = 0$
- Treat infinite values as a missing value
- Remove observations with 100% missing values
- Impute missing values with last non-missing observation

## Checklist Item 3.1 – Franchise power

**Formula**

$$fran_power_t = Percentile_{Cum}\left(\frac{roa_8_gmean_pct_t + roc_8_gmean_pct_t + fcft_pct_t + margin_max_t}{4}\right)$$

$$margin_max_t = max[gms_pct_t, gm_8_gmean_pct_t]$$

$$roa_8_gmean_pct_t = Percentile_{cum}(roa_8_gmean)$$

$$roa_8_gmean = \left(\left(\prod_{t=0}^{-7} 1 + roa_t\right)^{\frac{1}{8}}\right) - 1$$

$$roa_t = \frac{ibcom_t}{at_t}$$

$$roc_8_gmean_pct_t = Percentile_{cum}(roc_8_gmean)$$

$$roc_8_gmean = \left(\left(\prod_{t=0}^{-7} 1 + roc_t\right)^{\frac{1}{8}}\right) - 1$$

$$roc_t = \frac{ebit_t}{capital_t}$$

$$capital_t = ppent_t + act_t - lct_t - ch_t$$

$$fcfa_pct_t = Percentile_{cum}\left(\frac{fcf_8y_t}{at_t}\right)$$

$$fcf_8y_t = \sum_{0}^{-7}(ni_t + dp_t - wcapch_t - capx_t)$$

$$gm_8_gmean_pct_t = \ Percentile_{cum}(gm_8_gmean)$$

$$gm_8_gmean = \left(\left(\prod_{t=0}^{-7} 1 + gm_t\right)^{\frac{1}{8}}\right) - 1$$

$$gm_t = \frac{(sale_t - cogs_t)}{sale_t}$$

$$gms_pct_t = \ Percentile_{cum}(gms)$$

$$gms_t = \frac{gm_8_gmean_avg_t}{gm_8_gmean_sd_t}$$

$$gm_8_gmean_avg_t = \frac{\sum_0^{-7}(gm_t)}{8}$$

$$gm_8_gmean_sd_t = \ std\left(\sum_{0}^{-7}(gm_t)\right)$$

*Where:*

$$Percentile_{Cum} = Cumulative\ percentile\ of\ value$$

$$wcapch_t = (act_t - lct_t) - (act_{t-1} - lct_{t-1})$$

**Implementation Notes – roa**

- Return on assets ($roa_t$) is defined as $ibcom_t/at_t$. When $ibcom_t$ is unavailable, set $roa_t = niadj_t/at_t$. When $niadj_t$ is unavailable, set $roa_t = ni_t/at_t$.
- When $ibcom_t = 0$ (or either $niadj_t = 0$ or $ni_t = 0$ (as applicable)), set $roa_t = 0$
- When $at_t < 0$, set $roa_t = 0$.
- Treat infinite values as a missing value
- Remove observations with 100% missing values
- Impute missing values with last non-missing observation

**Implementation Notes – roa_8_gmean**

- Shift values for [$roa_t \ldots roa_{t-7}$] to ensure positive values for geometric mean calculation.
- Perform the following steps on data subsets grouped by LPERMNO:
  - When the minimum value for [$roa_t \ldots roa_{t-7}$] < 0, add the absolute value of min([$roa_t \ldots roa_{t-7}$]) to [$roa_t \ldots roa_{t-7}$].
  - When the minimum value for [$roa_t \ldots roa_{t-7}$] > 0, add 0.1 to all values in [$roa_t \ldots roa_{t-7}$].
  - For both cases, add 1 to the final adjusted values in [$roa_t \ldots roa_{t-7}$].

- When calculating $roa_8_gmean_t$, subtract the shift value -1 from $roa_8_gmean_t$.
- After calculating $roa_8_geman_t$, revert $[roa_t \ldots roa_{t-7}]$ to original values.

**Implementation Notes – roc**

- When $ebit_t = 0$, set $roc_t = 0$
- When $capital_t < 0$, set $roc_t = 0$
- Treat infinite values as a missing value
- Remove observations with 100% missing values
- Impute missing values with last non-missing observation

**Implementation Notes – roc_8_gmean**

- Shift values for $[roc_t \ldots roc_{t-7}]$ to ensure positive values for geometric mean calculation.
- Perform the following steps on data subsets grouped by LPERMNO.
    - When the minimum value for $[roc_t \ldots roc_{t-7}] < 0$, add the absolute value of $\min([roc_t \ldots roc_{t-7}])$ to $[roc_t \ldots roc_{t-7}]$.
    - When the minimum value for $[roc_t \ldots roc_{t-7}] > 0$, add 0.1 to all values in $[roc_t \ldots roc_{t-7}]$.
    - For both cases, add 1 to the final adjusted values in $[roc_t \ldots roc_{t-7}]$.
    - When calculating $roc_8_gmean_t$, subtract the shift value -1 from $roc_8_gmean_t$.
    - After calculating $roc_8_geman_t$, revert $[roc_t \ldots roc_{t-7}]$ to original values.

**Implementation Notes – gm**

- When $sale_t = 0$, set $gm_t = 0$
- Treat infinite values as a missing value.
- Remove observations with 100% missing values.
- Impute missing values with last non-missing observation.

**Implementation Notes – gm_8_gmean**

- Shift values for $[gm_t \ldots gm_{t-7}]$ to ensure positive values for geometric mean calculation. Perform the following steps on data subsets grouped by LPERMNO.
- When the minimum value for $[gm_t \ldots gm_{t-7}] < 0$, add the absolute value of $\min([gm_t \ldots gm_{t-7}])$ to $[gm_t \ldots gm_{t-7}]$.
- When the minimum value for $[gm_t \ldots gm_{t-7}] > 0$, add 0.1 to all values in $[gm_t \ldots gm_{t-7}]$. For both cases, add 1 to the final adjusted values in $[gm_t \ldots gm_{t-7}]$.
- When calculating $gm_8_gmean_t$, subtract the shift value -1 from $gm_8_gmean_t$.
- After calculating $gm_8_geman_t$, revert $[gm_t \ldots gm_{t-7}]$ to original values.

**Implementation Notes – gms**

- When $gm_8_gmean_avg_t = 0$ and $gm_8_gmean_sd_t$, set $gms_t = 0$
- Treat infinite values as a missing value.
- Remove observations with 100% missing values.
- Impute missing values with last non-missing observation.

## Checklist Item 3.2.1 – Current Profitability

**Formula**

$$fs_accrual_t = \begin{cases} 1 \text{ if } accrual_t > 0 \\ 0 \text{ otherwise} \end{cases}$$

$$fs_fcfta_t = \begin{cases} 1 \text{ if } fcfta_t > 0 \\ 0 \text{ otherwise} \end{cases}$$

$$accrual_t = fcfta_t - roa_t$$

$$fcfta_t = fcf_t - at_t$$

$$fcf_t = ni_t + dp_t - wcapch_t - capx_t$$

$$fs_roa_t = \begin{cases} 1 \text{ if } roa_t > 0 \\ 0 \text{ otherwise} \end{cases}$$

$$roa_t = \frac{ibcom_t}{at_t}$$

$Where$:
$$wcapch_t = (act_t - lct_t) - (act_{t-1} - lct_{t-1})$$

**Implementation Notes**
- Treat infinite values as a missing value.
- Remove observations with 100% missing values.
- Impute missing values with last non-missing observation.

## Checklist Item 3.2.2 – Financial Stability

**Formula**

$$lever_t = \left(\frac{dltt_{t-1}}{at_{t-1}}\right) - \left(\frac{dltt_t}{at_t}\right)$$

$$fs_lever_t = \begin{cases} 1 \text{ if } lever_t > 0 \\ 0 \text{ otherwise} \end{cases}$$

$$liquid_t = (current_t - current_{t-1})$$

$$fs_liquid_t = \begin{cases} 1 \text{ if } liquid_t > 0 \\ 0 \text{ otherwise} \end{cases}$$

$$current_t = \left(\frac{act_t}{lct_t}\right)$$

$$neqiss_t = \frac{(shrout_t - shrout_{t-1}) * \left(\frac{prc_t}{cfacpr_t} + \frac{prc_{t-1}}{cfacpr_{t-1}}\right)}{2}$$

$$fs_neqiss_t = \begin{cases} 1 \text{ if } neqiss_t > 0 \\ 0 \text{ otherwise} \end{cases}$$

**Implementation Notes – all**
- Treat infinite values as a missing value.
- Remove observations with 100% missing values.
- Impute missing values with last non-missing observation.

**Implementation Notes – lever**

- When $dltt_{t-1}$  0 and $at_{t-1} = 0$ and $dltt_t \neq 0$ and $at_t \neq 0$, set $lever_t = 0 - (dltt_t / at_t)$

**Implementation Notes – current**

- When $lct_t = 0$ and $act_t \neq 0$, set $current_t = act_t + 1$
- When $lct_t = 0$ and $act_t = 0$, set $current_t = 1$
- When $lct_{t-1} = 0$ and $act_{t-1} \neq 0$, set $current_{t-1} = act_{t-1} + 1$
- When $lct_{t-1} = 0$ and $act_{t-1} = 0$, set $current_{t-1} = 1$
- Treat infinite values as a missing value.
- Remove observations with 100% missing values.
- Impute missing values with last non-missing observation.

## Checklist Item 3.2.3 – Recent Operational Improvements

**Formula**

$$roa_change_t = roa_t - roa_{t-1}$$

$$fs_roa_t = \begin{cases} 1 \text{ if } roa_change_t > 0 \\ 0 \text{ otherwise} \end{cases}$$

$$fcfta_change_t = fcfta_t - fcfta_{t-1}$$

$$fs_fcfta_change_t = \begin{cases} 1 \text{ if } fcfta_change_t > 0 \\ 0 \text{ otherwise} \end{cases}$$

$$gm_change_t = gm_t - gm_{t-1}$$

$$fs_gm_t = \begin{cases} 1 \text{ if } gm_change_t > 0 \\ 0 \text{ otherwise} \end{cases}$$

$$turnover_change_t = \left(\frac{sale_t}{at_{t-1}}\right) - \left(\frac{sale_{t-1}}{at_{t-2}}\right)$$

$$fs_turnover_t = \begin{cases} 1 \text{ if } turnover_change_t > 0 \\ 0 \text{ otherwise} \end{cases}$$

**Implementation Notes**

- See previous implementation notes for $roa_t$, $fcfta_t$, $gm_t$
- Treat infinite values as a missing value.
- Remove observations with 100% missing values.
- Impute missing values with last non-missing observation.

## Checklist Item 3.2.4 – Financial Strength

**Formula**

$$fin_strength_t = (fs_roa_t + fs_fcfta_t + fs_accrual_t + fs_lever_t + fs_liquid_t + fs_neqiss_t + fs_roa_t + fs_fcfta_change_t + fs_gm_t + fs_turnover_t)/10$$

## Checklist Item 3.2.5 – Identify Quality

**Formula**

$$quality_t = \left(\frac{fran_power_t}{2}\right) + \left(\frac{fin_strength_t}{2}\right)$$

## Appendix 7 – Buffett & Clark Variables Overview

| Checklist | Variable |
|---|---|
| 1. Does the business have an identifiable consumer monopoly? | None. Defined by all other checklist items. |
| 2. Are the earnings of the company strong and showing an upward trend? | epsfi_8_gmean_pct |
| 3. Is the company conservatively financed? | fin_con1_8_gmean_pct<br>fin_con2_8_gmean_pct<br>fin_con3_8_gmean_pct |
| 4. Does the business consistently earn a high rate of return on shareholders' equity? | roe_8_gmean_pct |
| 5. Does the business get to retain its earnings? | re_use_8_gmean_pct |
| 6. How much does the business have to spend on maintaining current operations? | re_maintain_8_gmean_pct |
| 7. Is the company free to reinvest retained earnings in new business opportunities, expansion of operations, or share repurchases? | re_perc_8_gmean<br>dvt_perc_8_gmean |
| a. How good a job does the management do at this? | re_return_pct |
| b. Are share repurchases driving eps growth? | neqiss_8_gmean_pct |
| 8. Is the company free to adjust prices to inflation? | rev_inf_8_gmean_pct |
| 9. Will the value added by retained earnings increase the market value of the company? | See #7 |
| **Price Metrics** | **Variable** |
| 10. Expected CAGR (eps) | price_cagr |
| 11. Expected CAGR (roe) | tot_cagr |
| 12. Target price for 20% CAGR | price_roe_pv<br>price_roe_ms |
| 13. Valuation relative to 10 year US Treasury Note | price_tsy_implied<br>price_tsy_implied_ms |

## Appendix 8 – Buffett & Clark Variable Details

Note: Unless otherwise noted, variable names refer to variables names used on the Wharton Research Data Services (WRDS) web query form for the CCM database.

| Checklist Item 2 – Are the earnings of the company strong and showing an upward trend? |
|---|

**Formula**

$$epsfi_8_cagr_pct = Percentile_{Cum}\left(\left(\left(\prod_{t=0}^{-7} 1 + epsfi_t\right)^{\frac{1}{8}}\right) - 1\right)$$

*Where:*

$Percentile_{Cum} = Cumulative\ percentile\ of\ value$

**Implementation Notes – epsfi_8_gmean**

- Shift values for [$epsfi_t$…$epsfi_{t-7}$] to ensure positive values for geometric mean calculation.
- Perform the following steps on data subsets grouped by LPERMNO:
  - When the minimum value for [$epsfi_t$…$epsfi_{t-7}$] < 0, add the absolute value of min([$epsfi_t$…$epsfi_{t-7}$]) to [$epsfi_t$…$epsfi_{t-7}$].
  - When the minimum value for [$epsfi_t$…$epsfi_{t-7}$] > 0, add 0.1 to all values in [$epsfi_t$…$epsfi_{t-7}$]. For both cases, add 1 to the final adjusted values in [$epsfi_t$…$epsfi_{t-7}$].
  - When calculating $epsfi_8_gmean_t$, subtract the shift value -1 from $epsfi_8_gmean_t$.
  - After calculating $epsfi_8_geman_t$, revert [$epsfi_t$… $epsfi_{t-7}$] to original values.

| Checklist Item 3 – Is the Company Conservatively Financed (#1) |
|---|

**Formula**

$$fin_con1_8_gmean_pct = Percentile_{Cum}\left(\left(\left(\prod_{t=0}^{-7} 1 + fin_con1_t\right)^{\frac{1}{8}}\right) - 1\right)$$

*Where*:

$$fin_con1 = \frac{dltt}{niadj}$$

**Implementation Notes – fin_con1**

- When $niadj_t$ is missing, set fin_con1 = $dltt_t$ / $ni_t$
- When $niadj_t$ is 0, set fin_con1 = $dltt_t$
- Treat infinite values as a missing value.
- Remove observations with 100% missing values.
- Impute missing values with last non-missing observation.

**Implementation Notes – fin_con1_8_gmean**

- Shift values for [$fin_con1_t$…$fin_con1_{t-7}$] to ensure positive values for geometric mean calculation.
- Perform the following steps on data subsets grouped by LPERMNO:
  - When the minimum value for [$fin_con1_t$…$fin_con1_{t-7}$] < 0, add the absolute value of

- min([$fin_con1_t \ldots fin_con1_{t-7}$]) to [$fin_con1_t \ldots fin_con1_{t-7}$].
- When the minimum value for [$fin_con1_t \ldots fin_con1_{t-7}$] > 0, add 0.1 to all values in [$fin_con1_t \ldots fin_con1_{t-7}$].
- For both cases, add 1 to the final adjusted values in [$fin_con1_t \ldots fin_con1_{t-7}$].
- When calculating $fin_con1_8_gmean_t$, subtract the shift value -1 from $fin_con1_8_gmean_t$.
- After calculating $fin_con1_8_geman_t$, revert [$fin_con1_t \ldots fin_con1_{t-7}$] to original values.

## Checklist Item 3 – Is the Company Conservatively Financed (#2)

**Formula**

$$fin_con2_8_gmean_pct = Percentile_{Cum}\left(\left(\left(\prod_{t=0}^{-7} 1 + fin_con2_t\right)^{\frac{1}{8}}\right) - 1\right)$$

*Where*:

$$fin_con2 = \frac{dltt}{seq}$$

**Implementation Notes – fin_con2**

- When $seq_t$ is missing, set $fin_con2_t = dltt / (ceq_t + pstk_t)$
- When $seq_t$ is missing and $ceq_t$ or $pstk_t$ is missing, set $fin_con2_t = dltt / (at_t - lt_t)$
- When $seq_t$ is missing and $ceq_t$ or $pstk_t$ is missing and $at_t$ or $lt_t$ is missing, set fin_con2 = 0
- Treat infinite values as a missing value.
- Remove observations with 100% missing values.
- Impute missing values with last non-missing observation.

**Implementation Notes – fin_con2_8_gmean**

- Shift values for [$fin_con2_t \ldots fin_con2_{t-7}$] to ensure positive values for geometric mean calculation.
- Perform the following steps on data subsets grouped by LPERMNO:
  - When the minimum value for [$fin_con2_t \ldots fin_con2_{t-7}$] < 0, add the absolute value of min([$fin_con2_t \ldots fin_con2_{t-7}$]) to [$fin_con2_t \ldots fin_con2_{t-7}$].
  - When the minimum value for [$fin_con2_t \ldots fin_con2_{t-7}$] > 0, add 0.1 to all values in [$fin_con2_t \ldots fin_con2_{t-7}$].
  - For both cases, add 1 to the final adjusted values in [$fin_con2_t \ldots fin_con2_{t-7}$].
  - When calculating $fin_con2_8_gmean_t$, subtract the shift value -1 from $fin_con2_8_gmean_t$.
  - After calculating $fin_con2_8_geman_t$, revert [$fin_con2_t \ldots fin_con2_{t-7}$] to original values.

## Checklist Item 3 – Is the Company Conservatively Financed (#3)

**Formula**

$$fin_con3_8_gmean_pct = Percentile_{Cum}\left(\left(\left(\prod_{t=0}^{-7} 1 + fin_con3_t\right)^{\frac{1}{8}}\right) - 1\right)$$

*Where*:

$$fin_con3 = \frac{act}{lt}$$

**Implementation Notes – fin_con3**

- When $act_t > lt_t$ and $lt_t < 1$, set $fin_con3_t = act_t$
- When $act_t = 0$ and $lt_t = 0$, set $fin_con3_t = 1$
- Treat infinite values as a missing value.
- Remove observations with 100% missing values.
- Impute missing values with last non-missing observation.

**Implementation Notes – fin_con3_8_gmean**

- Shift values for [$fin_con3_t \ldots fin_con3_{t-7}$] to ensure positive values for geometric mean calculation.
- Perform the following steps on data subsets grouped by LPERMNO:
  - When the minimum value for [$fin_con3_t \ldots fin_con3_{t-7}$] < 0, add the absolute value of min([$fin_con3_t \ldots fin_con3_{t-7}$]) to [$fin_con3_t \ldots fin_con3_{t-7}$].
  - When the minimum value for [$fin_con3_t \ldots fin_con3_{t-7}$] > 0, add 0.1 to all values in [$fin_con3_t \ldots fin_con3_{t-7}$].
  - For both cases, add 1 to the final adjusted values in [$fin_con3_t \ldots fin_con3_{t-7}$].
  - When calculating $fin_con3_8_gmean_t$, subtract the shift value -1 from $fin_con3_8_gmean_t$.
  - After calculating $fin_con3_8_geman_t$, revert [$fin_con3_t \ldots fin_con3_{t-7}$] to original values.

**Checklist Item 4 – Does the business consistently earn a high rate of return on shareholders' equity?**

**Formula**

$$roe_8_gmean_pct = Percentile_{Cum}\left(\left(\left(\prod_{t=0}^{-7} 1 + roe_t\right)^{\frac{1}{8}}\right) - 1\right)$$

$Where$:

$$roe = \frac{niadj}{seq}$$

**Implementation Notes – roe**

- When $niadj_t$ is missing, set $roe_t = ni_t / seq_t$
- When $seq_t$ is missing, set $roe_t = niadj_t / (ceq_t + pstk_t)$
- When $seq_t$ is missing and $ceq_t$ or $pstk_t$ is missing, set $roe_t = niadj_t / (at_t – lt_t)$
- Treat infinite values as a missing value.
- Remove observations with 100% missing values.
- Impute missing values with last non-missing observation.

**Implementation Notes – roe_8_gmean**

- Shift values for [$roe_t \ldots roe_{t-7}$] to ensure positive values for geometric mean calculation.
- Perform the following steps on data subsets grouped by LPERMNO:
  - When the minimum value for [$roe_t \ldots roe_{t-7}$] < 0, add the absolute value of min([$roe_t \ldots roe_{t-7}$]) to [$roe_t \ldots roe_{t-7}$].
  - When the minimum value for [$roe_t \ldots roe_{t-7}$] > 0, add 0.1 to all values in [$roe_t \ldots roe_{t-7}$].
  - For both cases, add 1 to the final adjusted values in [$roe_t \ldots roe_{t-7}$].
  - When calculating $roe_8_gmean_t$, subtract the shift value -1 from $roe_8_gmean_t$.
  - After calculating $roe_8_geman_t$, revert [$roe_t \ldots roe_{t-7}$] to original values.

| Checklist Item 5 – Does the business get to retain its earnings? |
| --- |

**Formula**

$$re_use_8_gmean_pct = Percentile_{Cum}\left(\left(\left(\prod_{t=0}^{-7} 1 + re_use_t\right)^{\frac{1}{8}}\right) - 1\right)$$

$Where$:

$$re_use = \frac{re}{niadj}$$

**Implementation Notes – re_use**

- When $niadj_t$ is missing, set $re_use_t = re_t / ni_t$
- When $niadj_t = 0$, set $re_use_t = re_t$
- When $niadj_t$ is missing and $ni_t = 0$, set $re_use_t = re_t$
- When $re_t = 0$ and $niadj_t = 0$ and $ni_t = 0$, set $re_use_t = 0$
- Treat infinite values as a missing value.
- Remove observations with 100% missing values.
- Impute missing values with last non-missing observation.

**Implementation Notes – re_use_8_gmean**

- Shift values for $[re_use_t \ldots re_use_{t-7}]$ to ensure positive values for geometric mean calculation.
- Perform the following steps on data subsets grouped by LPERMNO:
  - When the minimum value for $[re_use_t \ldots re_use_{t-7}] < 0$, add the absolute value of $\min([re_use_t \ldots re_use_{t-7}])$ to $[re_use_t \ldots re_use_{t-7}]$.
  - When the minimum value for $[re_use_t \ldots re_use_{t-7}] > 0$, add 0.1 to all values in $[re_use_t \ldots re_use_{t-7}]$.
  - For both cases, add 1 to the final adjusted values in $[re_use_t \ldots re_use_{t-7}]$.
  - When calculating $re_use_8_gmean_t$, subtract the shift value -1 from $re_use_8_gmean_t$.
  - After calculating $re_use_8_geman_t$, revert $[re_use_t \ldots re_use_{t-7}]$ to original values.

| Checklist Item 6 – How much does the business have to spend on maintaining current operations? |
| --- |

**Formula**

$$re_maintain_8_gmean_pct = Percentile_{Cum}\left(\left(\left(\prod_{t=0}^{-7} 1 + re_maintain_t\right)^{\frac{1}{8}}\right) - 1\right)$$

$Where$:

$$re_maintain = 1 - \frac{capx}{re}$$

**Implementation Notes – re_maintain**

- When $capx_t > re_t$, set $re_maintain_t = abs(re_maintain_t)*-1$
- When $capx_t \neq 0$ and $re_t <= 0$, set $re_maintain_t = abs(cap_x_t)*-1$
- When $capx_t = 0$ and $re_t <= 0$, set $re_maintain_t = 1$
- Treat infinite values as a missing value.

- Remove observations with 100% missing values.
- Impute missing values with last non-missing observation.

**Implementation Notes – re_maintain_8_gmean**

- Shift values for [$re_maintain_t$…$re_maintain_{t-7}$] to ensure positive values for geometric mean calculation.
- Perform the following steps on data subsets grouped by LPERMNO:
  - When the minimum value for [$re_maintain_t$…$re_maintain_{t-7}$] < 0, add the absolute value of min([$re_maintain_t$…$re_maintain_{t-7}$]) to [$re_maintain_t$…$re_maintain_{t-7}$].
  - When the minimum value for [$re_maintain_t$…$re_maintain_{t-7}$] > 0, add 0.1 to all values in [$re_maintain_t$…$re_maintain_{t-7}$].
  - For both cases, add 1 to the final adjusted values in [$re_maintain_t$…$re_maintain_{t-7}$].
  - When calculating $re_maintain_8_gmean_t$, subtract the shift value -1 from $re_maintain_8_gmean_t$.
  - After calculating $re_maintain_8_geman_t$, revert [$re_maintain_t$… $re_maintain_{t-7}$] to original values.

Checklist Item 7 – Is the company free to reinvest retained earnings in new business opportunities, expansion of operations, or share repurchases?

**Formulas**

$$re_perc_8_gmean = \left(\left(\prod_{t=0}^{-7} 1 + re_perc_t\right)^{\frac{1}{8}}\right) - 1$$

$$dvt_perc_8_gmean = \left(\left(\prod_{t=0}^{-7} 1 + dvt_perc_t\right)^{\frac{1}{8}}\right) - 1$$

**Implementation Notes – re_perc**

- When $re_perc_t$ is missing, set $re_perc_t = 0$

**Implementation Notes – re_perc_8_gmean**

- Shift values for [$re_perc_t$…$re_perc_{t-7}$] to ensure positive values for geometric mean calculation.
- Perform the following steps on data subsets grouped by LPERMNO:
  - When the minimum value for [$re_perc_t$…$re_perc_{t-7}$] < 0, add the absolute value of min([$re_perc_t$…$re_perc_{t-7}$]) to [$re_perc_t$…$re_perc_{t-7}$].
  - When the minimum value for [$re_perc_t$…$re_perc_{t-7}$] > 0, add 0.1 to all values in [$re_perc_t$…$re_perc_{t-7}$].
  - For both cases, add 1 to the final adjusted values in [$re_perc_t$…$re_perc_{t-7}$].
  - When calculating $re_perc_8_gmean_t$, subtract the shift value -1 from $re_perc_8_gmean_t$.
  - After calculating $re_perc_8_geman_t$, revert [$re_perc_t$… $re_perc_{t-7}$] to original values.

**Implementation Notes – dvt_perc**

- When $dvt_perc_t > 1$, set $dvt_perc_t = 1$
- When $dvt_perc_t > 1$ and $niadj_t = 0$, set $dvt_perc_t = 1$
- When $dvt_perc_t = 0$ and $niadj_t = 0$, set $dvt_perc_t = 0$

**Implementation Notes – dvt_perc_8_gmean**

- Shift values for [$dvt_perc_t \ldots dvt_perc_{t-7}$] to ensure positive values for geometric mean calculation.
- Perform the following steps on data subsets grouped by LPERMNO:
  - When the minimum value for [$dvt_perc_t \ldots dvt_perc_{t-7}$] < 0, add the absolute value of min([$dvt_perc_t \ldots dvt_perc_{t-7}$]) to [$dvt_perc_t \ldots dvt_perc_{t-7}$].
  - When the minimum value for [$dvt_perc_t \ldots dvt_perc_{t-7}$] > 0, add 0.1 to all values in [$dvt_perc_t \ldots dvt_perc_{t-7}$].
  - For both cases, add 1 to the final adjusted values in [$dvt_perc_t \ldots dvt_perc_{t-7}$].
  - When calculating $dvt_perc_8_gmean_t$, subtract the shift value -1 from $dvt_perc_8_gmean_t$.
  - After calculating $dvt_perc_8_geman_t$, revert [$dvt_perc_t \ldots dvt_perc_{t-7}$] to original values.

## Checklist Item 7a – How good a job does the management do at this?

**Formula**

$$re_return_pe = Percentile_{Cum}(re_return)$$

*Where*:

$$re_return = \frac{mean\,[re_t \ldots re_{t-2}] - mean\,[re_{t-5} \ldots re_{t-7}]}{\sum_{t=0}^{-7} re_t}$$

## Checklist Item 7b – Are share repurchases driving earnings per share growth?

**Formula**

$$neqiss_8_gmean_pct = Percentile_{Cum}\left(\left(\left(\prod_{t=0}^{-7} 1 + neqiss_t\right)^{\frac{1}{8}}\right) - 1\right)$$

*Where*:

$$neqiss = \frac{\left((shrout_t * cfacshr_t) - (shrout_{t-1} * cfacshr_{t-1})\right) * \left(\frac{prc_t}{cfacpr_t}\right) + \left(\frac{prc_{t-1}}{cfacpr_{t-1}}\right)}{2}$$

**Implementation notes – neqiss**

- When $prc_t$ is missing, set $prc_t = 0$. Note: although observations with missing prices are removed from the dataset, lagged prices are required for calculating neqiss.
- Treat infinite values as a missing value.
- Remove observations with 100% missing values.
- Impute missing values with last non-missing observation.

**Implementation notes – neqiss_8_gmean**

- Shift values for [$neqiss_t \ldots neqiss_{t-7}$] to ensure positive values for geometric mean calculation.
- Perform the following steps on data subsets grouped by LPERMNO:
  - When the minimum value for [$neqiss_t \ldots neqiss_{t-7}$] < 0, add the absolute value of min([$neqiss_t \ldots neqiss_{t-7}$]) to [$neqiss_t \ldots neqiss_{t-7}$].
  - When the minimum value for [$neqiss_t \ldots neqiss_{t-7}$] > 0, add 0.1 to all values in [$neqiss_t \ldots neqiss_{t-7}$].
  - For both cases, add 1 to the final adjusted values in [$neqiss_t \ldots neqiss_{t-7}$].
  - When calculating $neqiss_8_gmean_t$, subtract the shift value -1 from $neqiss_8_gmean_t$. After calculating $neqiss_8_geman_t$, revert [$neqiss_t \ldots neqiss_{t-7}$] to original values.

## Checklist Item 8 – Is the company free to adjust prices to inflation?

**Formula**

$$rev_inf_8_gmean_pct = Percentile_{Cum}\left(\left(\left(\prod_{t=0}^{-7} 1 + rev_inf_t\right)^{\frac{1}{8}}\right) - 1\right)$$

$Where$:

$$rev_inf = \left(\left(\frac{sale_t}{sale_{t-1}}\right) - 1\right) - cpiind_yoy$$

$$cpiind_yoy = \left(\left(\frac{cpiind_t}{cpiind_{t-12}}\right) - 1\right)$$

**Implementation notes – rev_inf**

- When $sale_{t-1} = 0$, set $rev_inf_t = 0$
- When $sale_t = 0$ and $sale_{t-1} = 0$, set $rev_inf_t = 0$
- Treat infinite values as a missing value.
- Remove observations with 100% missing values.
- Impute missing values with last non-missing observation.

**Implementation notes – rev_inf_8_gmean**

- Shift values for $[rev_inf_t \ldots rev_inf_{t-7}]$ to ensure positive values for geometric mean calculation.
- Perform the following steps on data subsets grouped by LPERMNO:
  - When the minimum value for $[rev_inf_t \ldots rev_inf_{t-7}] < 0$, add the absolute value of $\min([rev_inf_t \ldots rev_inf_{t-7}])$ to $[rev_inf_t \ldots rev_inf_{t-7}]$.
  - When the minimum value for $[rev_inf_t \ldots rev_inf_{t-7}] > 0$, add 0.1 to all values in $[rev_inf_t \ldots rev_inf_{t-7}]$.
  - For both cases, add 1 to the final adjusted values in $[rev_inf_t \ldots rev_inf_{t-7}]$.
  - When calculating $rev_inf_8_gmean_t$, subtract the shift value -1 from $rev_inf_8_gmean_t$.
  - After calculating $rev_inf_8_geman_t$, revert $[rev_inf_t \ldots rev_inf_{t-7}]$ to original values.

## Checklist Item 10 – Price Variable: Expected CAGR (eps)

**Formula**

$$price_cagr = \left(\left(\frac{price_eps_future}{adj_close}\right)^{\frac{1}{10}}\right) - 1$$

$Where$:

$$price_eps_future = (epsfi_fv - epsfi) * sign(epsfi) + epsfi$$

$$epsfi_fv = sign(1 + epsfi_8_cagr) * sign(epsfi) * abs\big(epsfi * ((1 + epsfi_8_cagr)^{10})\big)$$

$$epsfi_8_cagr = \left(\left(\prod_{t=0}^{-7} 1 + epsfi_t\right)^{\frac{1}{8}}\right) - 1$$

**Implementation Notes**

- When $price_eps_future_t < 0$, set $price_cagr_t = -1$

## Checklist Item 11 – Price Variable: Expected CAGR (roe)

$$tot_cagr = \left(\left(\frac{adj_close_fv \ + \ dvt_fv}{adj_close}\right)^{\frac{1}{10}}\right) - 1$$

$Where:$

$$adj_close_fv = \ eps_fv * min\ [pe_{t-8} \dots pe_t]$$

$$dvt_fv = \sum_{t=1}^{10} \big((bkvlps_fv - bkvlps) * sign(bkvlps * dvt_perc_8_gmean) \ + \ bkvlps\big)$$

$$\begin{aligned} eps_fv = roe_8_gmean \\ &* \Big(\big(sign(1 + re_perc_8_gmean\ ) * \ sign(bkvlps * dvt_perc_8_gmean) \\ &* abs\big(bkvlps * ((1 + re_perc_8_gmean)^{10})\big)\big) * sign(bkvlps * dvt_perc_8_gmean) \\ &+ bkvlps\Big) \end{aligned}$$

$$\begin{aligned} bkvlps_fv = sign(1 + re_perc_8_gmean) * sign(bkvlps * dvt_perc_8_gmean) \\ &* abs\big(bkvlps * dvt_perc_8_gmean * ((1 + re_perc_8_gmean)^{10})\big) \end{aligned}$$

$$roe_8_gmean = \left(\left(\prod_{t=0}^{-7} 1 + roe_t\right)^{\frac{1}{8}}\right) - 1$$

$$re_perc = \left(1 - \frac{dvt}{niadj}\right) * roe$$

$$dvt_perc = \left(\frac{dvt}{niadj}\right) - roe$$

$$pe = \frac{adj_close}{epsfi}$$

**Implementation Notes – dvt_fv**

- When $bkvlps_t <= 0$, set $dvt_fv_t = 0$

**Implementation Notes – tot_cagr**

- When $adj_close_fv_t <=0$, set $tot_cagr_t = -1$

## Checklist Item 12 – Target price for 20% CAGR

**Formula**

$$price_roe_pv = \frac{bkvlps_fv}{(1 + 20\%)^{10}}$$

$$price_roe_ms = \frac{adj_close}{price_roe_pv}$$

**Implementation Notes – price_roe_ms**

- When $price_roe_pv_t = 0$, set $price_roe_ms_t = adj_close_t + 1$
- When $price_roe_ms_t <= 0$, set $price_roe_ms_t = (0.01 + adj_close_t + abs(price_roe_pv_t)) / 0.01$
- When $price_roe_ms_t$ is infinite, set $price_roe_ms_t = 0$

**Implementation Notes – bkvlps_fv**

- When $re_perc_8_gmean_t < -1$, set $bkvlps_fv_t = \text{sign}(1 + \text{re_perc_8_gmean}) * \text{sign}(\text{bkvlps} * \text{dvt_perc_8_gmean}) * \text{abs}(\text{bkvlps} * \text{dvt_perc_8_gmean})$

## Checklist Item 13 – Valuation relative to 10 year US Treasury Note

**Formula**

$$price_tsy_implied = \frac{epsfi}{tsy_10_yield}$$

$$price_tsy_implied_ms = \frac{adj_close}{price_tsy_implied}$$

**Implementation Notes – price_tsy_implied_ms**

- When $price_tsy_implied_t = 0$, set $price_tsy_implied_ms_t = adj_close_t + 1$

**Appendix 9 – Monthly Return Data (Post-Fees) for Model 11 and Model 12**

Note: Monthly return data for Model 12 is not available until 6/29/1990 due to the use of a 10-year alpha response variable.

| Date | Model 11 | Model 12 |
|---|---|---|
| 6/28/1985 | 0.000 | -- |
| 7/31/1985 | 0.020 | -- |
| 8/30/1985 | -0.013 | -- |
| 9/30/1985 | -0.033 | -- |
| 10/31/1985 | 0.042 | -- |
| 11/29/1985 | 0.072 | -- |
| 12/31/1985 | 0.036 | -- |
| 1/31/1986 | -0.003 | -- |
| 2/28/1986 | 0.074 | -- |
| 3/31/1986 | 0.065 | -- |
| 4/30/1986 | -0.001 | -- |
| 5/30/1986 | 0.081 | -- |
| 6/30/1986 | -0.004 | -- |
| 7/31/1986 | -0.087 | -- |
| 8/29/1986 | 0.097 | -- |
| 9/30/1986 | -0.045 | -- |
| 10/31/1986 | 0.057 | -- |
| 11/28/1986 | 0.033 | -- |
| 12/31/1986 | -0.027 | -- |
| 1/30/1987 | 0.159 | -- |
| 2/27/1987 | 0.039 | -- |
| 3/31/1987 | 0.045 | -- |
| 4/30/1987 | 0.005 | -- |
| 5/29/1987 | 0.021 | -- |
| 6/30/1987 | -0.003 | -- |
| 7/31/1987 | 0.079 | -- |
| 8/31/1987 | 0.010 | -- |
| 9/30/1987 | -0.025 | -- |
| 10/30/1987 | -0.268 | -- |
| 11/30/1987 | -0.044 | -- |
| 12/31/1987 | 0.082 | -- |
| 1/29/1988 | 0.035 | -- |
| 2/29/1988 | 0.071 | -- |
| 3/31/1988 | 0.000 | -- |
| 4/29/1988 | 0.017 | -- |
| 5/31/1988 | -0.015 | -- |
| 6/30/1988 | -0.003 | -- |
| 7/29/1988 | 0.001 | -- |
| 8/31/1988 | -0.022 | -- |
| 9/30/1988 | 0.052 | -- |
| 10/31/1988 | 0.024 | -- |
| 11/30/1988 | -0.018 | -- |
| 12/30/1988 | 0.027 | -- |
| 1/31/1989 | 0.066 | -- |
| 2/28/1989 | -0.011 | -- |
| 3/31/1989 | 0.027 | -- |
| 4/28/1989 | 0.057 | -- |
| 5/31/1989 | 0.033 | -- |
| 6/30/1989 | -0.003 | -- |

| Date | Model 11 | Model 12 |
|---|---|---|
| 7/31/1989 | 0.120 | -- |
| 8/31/1989 | 0.033 | -- |
| 9/29/1989 | -0.013 | -- |
| 10/31/1989 | -0.035 | -- |
| 11/30/1989 | 0.018 | -- |
| 12/29/1989 | 0.027 | -- |
| 1/31/1990 | -0.054 | -- |
| 2/28/1990 | 0.010 | -- |
| 3/30/1990 | 0.028 | -- |
| 4/30/1990 | -0.032 | -- |
| 5/31/1990 | 0.098 | -- |
| 6/29/1990 | -0.003 | 0.000 |
| 7/31/1990 | -0.009 | -0.011 |
| 8/31/1990 | -0.105 | -0.081 |
| 9/28/1990 | -0.070 | -0.055 |
| 10/31/1990 | -0.005 | -0.010 |
| 11/30/1990 | 0.086 | 0.067 |
| 12/31/1990 | 0.035 | 0.024 |
| 1/31/1991 | 0.071 | 0.079 |
| 2/28/1991 | 0.072 | 0.063 |
| 3/28/1991 | 0.030 | 0.029 |
| 4/30/1991 | 0.017 | 0.022 |
| 5/31/1991 | 0.054 | 0.040 |
| 6/28/1991 | -0.002 | -0.005 |
| 7/31/1991 | 0.036 | 0.053 |
| 8/30/1991 | 0.026 | 0.021 |
| 9/30/1991 | -0.013 | 0.002 |
| 10/31/1991 | 0.020 | 0.044 |
| 11/29/1991 | -0.044 | -0.051 |
| 12/31/1991 | 0.123 | 0.101 |
| 1/31/1992 | 0.011 | 0.013 |
| 2/28/1992 | 0.028 | 0.026 |
| 3/31/1992 | -0.006 | -0.007 |
| 4/30/1992 | 0.016 | 0.017 |
| 5/29/1992 | 0.015 | -0.005 |
| 6/30/1992 | -0.002 | -0.003 |
| 7/31/1992 | 0.046 | 0.028 |
| 8/31/1992 | -0.011 | -0.019 |
| 9/30/1992 | 0.024 | 0.025 |
| 10/30/1992 | 0.026 | 0.020 |
| 11/30/1992 | 0.052 | 0.070 |
| 12/31/1992 | 0.005 | 0.008 |
| 1/29/1993 | 0.006 | 0.021 |
| 2/26/1993 | -0.014 | -0.009 |
| 3/31/1993 | 0.048 | 0.081 |
| 4/30/1993 | -0.038 | -0.054 |
| 5/28/1993 | 0.060 | 0.084 |
| 6/30/1993 | -0.002 | -0.003 |
| 7/30/1993 | -0.009 | 0.063 |

| Date | Model 11 | Model 12 |
|---|---|---|
| 8/31/1993 | 0.046 | 0.069 |
| 9/30/1993 | 0.005 | -0.020 |
| 10/29/1993 | 0.031 | -0.034 |
| 11/30/1993 | -0.016 | 0.087 |
| 12/31/1993 | 0.037 | 0.044 |
| 1/31/1994 | 0.014 | 0.042 |
| 2/28/1994 | 0.005 | 0.032 |
| 3/31/1994 | -0.038 | -0.057 |
| 4/29/1994 | 0.025 | 0.036 |
| 5/31/1994 | 0.009 | 0.024 |
| 6/30/1994 | -0.002 | -0.003 |
| 7/29/1994 | 0.016 | 0.028 |
| 8/31/1994 | 0.061 | 0.077 |
| 9/30/1994 | -0.019 | -0.019 |
| 10/31/1994 | 0.035 | -0.002 |
| 11/30/1994 | -0.021 | -0.026 |
| 12/30/1994 | 0.009 | 0.042 |
| 1/31/1995 | 0.024 | -0.003 |
| 2/28/1995 | 0.047 | 0.041 |
| 3/31/1995 | 0.040 | 0.058 |
| 4/28/1995 | 0.025 | 0.029 |
| 5/31/1995 | 0.040 | 0.061 |
| 6/30/1995 | -0.002 | -0.002 |
| 7/31/1995 | 0.034 | 0.058 |
| 8/31/1995 | -0.003 | -0.008 |
| 9/29/1995 | 0.017 | 0.018 |
| 10/31/1995 | -0.018 | -0.027 |
| 11/30/1995 | 0.046 | 0.084 |
| 12/29/1995 | 0.026 | 0.019 |
| 1/31/1996 | 0.023 | -0.002 |
| 2/29/1996 | 0.018 | 0.016 |
| 3/29/1996 | 0.040 | 0.041 |
| 4/30/1996 | 0.046 | 0.006 |
| 5/31/1996 | 0.030 | 0.062 |
| 6/28/1996 | -0.002 | -0.002 |
| 7/31/1996 | -0.076 | -0.063 |
| 8/30/1996 | 0.048 | 0.065 |
| 9/30/1996 | 0.071 | 0.069 |
| 10/31/1996 | 0.011 | 0.041 |
| 11/29/1996 | 0.075 | 0.108 |
| 12/31/1996 | 0.001 | 0.006 |
| 1/31/1997 | 0.024 | 0.039 |
| 2/28/1997 | 0.024 | 0.017 |
| 3/31/1997 | -0.015 | -0.010 |
| 4/30/1997 | 0.051 | 0.076 |
| 5/30/1997 | 0.073 | 0.097 |
| 6/30/1997 | -0.002 | -0.002 |
| 7/31/1997 | 0.084 | 0.065 |
| 8/29/1997 | 0.009 | -0.024 |

| Date | Model 11 | Model 12 |
|---|---|---|
| 9/30/1997 | 0.094 | 0.111 |
| 10/31/1997 | -0.015 | -0.049 |
| 11/28/1997 | 0.010 | 0.083 |
| 12/31/1997 | 0.008 | 0.008 |
| 1/30/1998 | 0.028 | -0.029 |
| 2/27/1998 | 0.098 | 0.157 |
| 3/31/1998 | 0.076 | 0.061 |
| 4/30/1998 | -0.006 | 0.007 |
| 5/29/1998 | -0.035 | -0.008 |
| 6/30/1998 | -0.002 | -0.002 |
| 7/31/1998 | -0.015 | 0.033 |
| 8/31/1998 | -0.154 | -0.127 |
| 9/30/1998 | 0.044 | 0.041 |
| 10/30/1998 | 0.116 | 0.055 |
| 11/30/1998 | 0.104 | 0.032 |
| 12/31/1998 | 0.092 | 0.137 |
| 1/29/1999 | 0.104 | 0.090 |
| 2/26/1999 | -0.053 | -0.043 |
| 3/31/1999 | 0.019 | -0.058 |
| 4/30/1999 | 0.057 | 0.066 |
| 5/28/1999 | -0.044 | -0.024 |
| 6/30/1999 | -0.002 | -0.002 |
| 7/30/1999 | -0.041 | 0.016 |
| 8/31/1999 | -0.039 | 0.004 |
| 9/30/1999 | -0.038 | -0.006 |
| 10/29/1999 | -0.015 | 0.048 |
| 11/30/1999 | 0.052 | 0.040 |
| 12/31/1999 | 0.047 | -0.023 |
| 1/31/2000 | -0.123 | -0.084 |
| 2/29/2000 | 0.027 | -0.056 |
| 3/31/2000 | 0.092 | 0.213 |
| 4/28/2000 | -0.009 | -0.042 |
| 5/31/2000 | 0.000 | -0.115 |
| 6/30/2000 | -0.002 | -0.002 |
| 7/31/2000 | 0.008 | 0.155 |
| 8/31/2000 | 0.061 | 0.113 |
| 9/29/2000 | 0.006 | 0.167 |
| 10/31/2000 | 0.028 | 0.164 |
| 11/30/2000 | 0.044 | -0.065 |
| 12/29/2000 | 0.216 | 0.185 |
| 1/31/2001 | 0.073 | -0.105 |
| 2/28/2001 | -0.001 | 0.050 |
| 3/30/2001 | -0.034 | -0.187 |
| 4/30/2001 | 0.062 | 0.060 |
| 5/31/2001 | 0.040 | -0.002 |
| 6/29/2001 | -0.002 | -0.002 |
| 7/31/2001 | 0.047 | 0.032 |
| 8/31/2001 | -0.016 | -0.172 |
| 9/28/2001 | -0.122 | -0.095 |
| 10/31/2001 | 0.114 | 0.111 |
| 11/30/2001 | 0.108 | 0.087 |
| 12/31/2001 | 0.057 | 0.452 |
| 1/31/2002 | 0.035 | -0.067 |
| 2/28/2002 | -0.027 | -0.209 |
| 3/28/2002 | 0.099 | 0.290 |

| Date | Model 11 | Model 12 |
|---|---|---|
| 4/30/2002 | 0.007 | -0.060 |
| 5/31/2002 | -0.022 | -0.040 |
| 6/28/2002 | -0.002 | -0.002 |
| 7/31/2002 | -0.109 | -0.121 |
| 8/30/2002 | -0.009 | 0.018 |
| 9/30/2002 | -0.101 | -0.118 |
| 10/31/2002 | 0.104 | 0.111 |
| 11/29/2002 | 0.126 | 0.212 |
| 12/31/2002 | -0.061 | -0.080 |
| 1/31/2003 | -0.021 | -0.011 |
| 2/28/2003 | -0.032 | 0.007 |
| 3/31/2003 | 0.017 | 0.006 |
| 4/30/2003 | 0.111 | 0.091 |
| 5/30/2003 | 0.072 | 0.113 |
| 6/30/2003 | -0.002 | -0.002 |
| 7/31/2003 | 0.022 | 0.042 |
| 8/29/2003 | 0.043 | 0.066 |
| 9/30/2003 | -0.021 | -0.020 |
| 10/31/2003 | 0.071 | 0.088 |
| 11/28/2003 | 0.034 | 0.054 |
| 12/31/2003 | 0.031 | 0.047 |
| 1/30/2004 | 0.019 | 0.036 |
| 2/27/2004 | 0.020 | 0.015 |
| 3/31/2004 | 0.017 | 0.015 |
| 4/30/2004 | -0.007 | -0.025 |
| 5/28/2004 | 0.008 | 0.023 |
| 6/30/2004 | -0.002 | -0.002 |
| 7/30/2004 | -0.053 | -0.044 |
| 8/31/2004 | -0.004 | 0.001 |
| 9/30/2004 | 0.020 | 0.025 |
| 10/29/2004 | 0.021 | 0.027 |
| 11/30/2004 | 0.053 | 0.058 |
| 12/31/2004 | 0.035 | 0.052 |
| 1/31/2005 | -0.024 | -0.016 |
| 2/28/2005 | 0.033 | 0.041 |
| 3/31/2005 | -0.014 | -0.021 |
| 4/29/2005 | -0.043 | -0.048 |
| 5/31/2005 | 0.045 | 0.022 |
| 6/30/2005 | -0.002 | -0.002 |
| 7/29/2005 | 0.057 | 0.039 |
| 8/31/2005 | -0.013 | -0.007 |
| 9/30/2005 | 0.010 | 0.021 |
| 10/31/2005 | -0.037 | -0.031 |
| 11/30/2005 | 0.042 | 0.046 |
| 12/30/2005 | 0.006 | 0.009 |
| 1/31/2006 | 0.046 | 0.029 |
| 2/28/2006 | -0.001 | 0.014 |
| 3/31/2006 | 0.024 | 0.038 |
| 4/28/2006 | 0.005 | 0.008 |
| 5/31/2006 | -0.037 | -0.047 |
| 6/30/2006 | -0.002 | -0.002 |
| 7/31/2006 | -0.021 | -0.014 |
| 8/31/2006 | 0.029 | 0.022 |
| 9/29/2006 | 0.018 | 0.034 |
| 10/31/2006 | 0.047 | 0.046 |

| Date | Model 11 | Model 12 |
|---|---|---|
| 11/30/2006 | 0.032 | 0.039 |
| 12/29/2006 | 0.002 | -0.009 |
| 1/31/2007 | 0.038 | 0.043 |
| 2/28/2007 | -0.004 | -0.015 |
| 3/30/2007 | 0.010 | 0.009 |
| 4/30/2007 | 0.042 | 0.023 |
| 5/31/2007 | 0.046 | 0.043 |
| 6/29/2007 | -0.002 | -0.002 |
| 7/31/2007 | -0.037 | -0.043 |
| 8/31/2007 | 0.004 | 0.003 |
| 9/28/2007 | 0.025 | 0.021 |
| 10/31/2007 | 0.021 | 0.029 |
| 11/30/2007 | -0.045 | -0.063 |
| 12/31/2007 | -0.008 | -0.025 |
| 1/31/2008 | -0.056 | -0.063 |
| 2/29/2008 | -0.007 | 0.011 |
| 3/31/2008 | 0.003 | -0.010 |
| 4/30/2008 | 0.057 | 0.066 |
| 5/30/2008 | 0.046 | 0.054 |
| 6/30/2008 | -0.002 | -0.002 |
| 7/31/2008 | -0.005 | -0.006 |
| 8/29/2008 | 0.037 | 0.044 |
| 9/30/2008 | -0.114 | -0.100 |
| 10/31/2008 | -0.240 | -0.230 |
| 11/28/2008 | -0.101 | -0.101 |
| 12/31/2008 | 0.053 | 0.048 |
| 1/30/2009 | -0.075 | -0.081 |
| 2/27/2009 | -0.118 | -0.131 |
| 3/31/2009 | 0.110 | 0.105 |
| 4/30/2009 | 0.185 | 0.192 |
| 5/29/2009 | 0.031 | 0.025 |
| 6/30/2009 | -0.002 | -0.002 |
| 7/31/2009 | 0.108 | 0.118 |
| 8/31/2009 | 0.043 | 0.037 |
| 9/30/2009 | 0.070 | 0.067 |
| 10/30/2009 | -0.034 | -0.035 |
| 11/30/2009 | 0.055 | 0.053 |
| 12/31/2009 | 0.054 | 0.052 |
| 1/29/2010 | -0.040 | -0.046 |
| 2/26/2010 | 0.058 | 0.054 |
| 3/31/2010 | 0.075 | 0.071 |
| 4/30/2010 | 0.036 | 0.039 |
| 5/28/2010 | -0.082 | -0.079 |
| 6/30/2010 | -0.002 | -0.002 |
| 7/30/2010 | 0.075 | 0.073 |
| 8/31/2010 | -0.050 | -0.061 |
| 9/30/2010 | 0.124 | 0.128 |
| 10/29/2010 | 0.041 | 0.048 |
| 11/30/2010 | 0.022 | 0.019 |
| 12/31/2010 | 0.075 | 0.075 |
| 1/31/2011 | 0.019 | 0.023 |
| 2/28/2011 | 0.048 | 0.041 |
| 3/31/2011 | 0.023 | 0.020 |
| 4/29/2011 | 0.027 | 0.037 |
| 5/31/2011 | -0.015 | -0.015 |

| Date | Model 11 | Model 12 |
|---|---|---|
| 6/30/2011 | -0.002 | -0.002 |
| 7/29/2011 | -0.039 | -0.040 |
| 8/31/2011 | -0.067 | -0.062 |
| 9/30/2011 | -0.106 | -0.096 |
| 10/31/2011 | 0.144 | 0.142 |
| 11/30/2011 | 0.007 | 0.009 |
| 12/30/2011 | -0.002 | -0.005 |
| 1/31/2012 | 0.061 | 0.053 |
| 2/29/2012 | 0.039 | 0.036 |
| 3/30/2012 | 0.014 | 0.021 |
| 4/30/2012 | -0.011 | -0.010 |
| 5/31/2012 | -0.073 | -0.072 |
| 6/29/2012 | -0.002 | -0.002 |
| 7/31/2012 | -0.003 | -0.004 |
| 8/31/2012 | 0.036 | 0.031 |
| 9/28/2012 | 0.022 | 0.020 |
| 10/31/2012 | -0.004 | -0.007 |
| 11/30/2012 | 0.022 | 0.023 |
| 12/31/2012 | 0.026 | 0.025 |
| 1/31/2013 | 0.081 | 0.074 |
| 2/28/2013 | 0.008 | 0.006 |
| 3/28/2013 | 0.043 | 0.038 |
| 4/30/2013 | -0.004 | 0.000 |
| 5/31/2013 | 0.037 | 0.034 |
| 6/28/2013 | -0.002 | -0.002 |
| 7/31/2013 | 0.068 | 0.064 |
| 8/30/2013 | -0.024 | -0.025 |
| 9/30/2013 | 0.056 | 0.051 |
| 10/31/2013 | 0.040 | 0.043 |
| 11/29/2013 | 0.027 | 0.030 |
| 12/31/2013 | 0.028 | 0.030 |
| 1/31/2014 | -0.046 | -0.045 |
| 2/28/2014 | 0.056 | 0.061 |
| 3/31/2014 | 0.003 | 0.004 |
| 4/30/2014 | -0.007 | -0.006 |
| 5/30/2014 | 0.015 | 0.017 |
| 6/30/2014 | -0.002 | -0.002 |
| 7/31/2014 | -0.035 | -0.030 |
| 8/29/2014 | 0.047 | 0.041 |
| 9/30/2014 | -0.049 | -0.041 |
| 10/31/2014 | 0.020 | 0.019 |
| 11/28/2014 | 0.038 | 0.030 |
| 12/31/2014 | 0.001 | 0.003 |
| 1/30/2015 | -0.025 | -0.021 |
| 2/27/2015 | 0.072 | 0.065 |
| 3/31/2015 | -0.007 | -0.010 |
| 4/30/2015 | -0.012 | -0.010 |
| 5/29/2015 | 0.019 | 0.017 |
| 6/30/2015 | -0.002 | -0.002 |
| 7/31/2015 | -0.006 | -0.024 |
| 8/31/2015 | -0.062 | -0.055 |
| 9/30/2015 | -0.058 | -0.064 |
| 10/30/2015 | 0.091 | 0.089 |
| 11/30/2015 | 0.000 | -0.002 |
| 12/31/2015 | -0.053 | -0.053 |

| Date | Model 11 | Model 12 |
|---|---|---|
| 1/29/2016 | -0.066 | -0.061 |
| 2/29/2016 | 0.023 | 0.037 |
| 3/31/2016 | 0.089 | 0.094 |
| 4/29/2016 | 0.015 | 0.037 |
| 5/31/2016 | 0.005 | -0.005 |
| 6/30/2016 | -0.002 | -0.002 |
| 7/29/2016 | 0.059 | 0.056 |
| 8/31/2016 | 0.001 | 0.004 |
| 9/30/2016 | 0.001 | 0.001 |
| 10/31/2016 | -0.039 | -0.033 |
| 11/30/2016 | 0.082 | 0.067 |
| 12/30/2016 | 0.007 | 0.003 |

**Bibliography**

Asness, C., Frazzini, A., Israel, R., & Moskowitz, T. (2015). Fact, fiction, and value investing. *The Journal of Portfolio Management*, *42*(1), 34-52. doi:10.2139/ssrn.2595747

Beneish, M. D. (1999). The detection of earnings manipulation. Financial Analysts Journal, 24-36. doi:10.2469/faj.v69.n2.1

Benjamini, Y., & Hochberg, Y. (1995). Controlling the false discovery rate: a practical and powerful approach to multiple testing. *Journal of the Royal Statistical Society. Series B (Methodological)*, 289-300. doi:10.1111/j.0016-7363.2006.00682.x

Boudoukh, J., Michaely, R., Richardson, M., & Roberts, M. R. (2007). On the importance of measuring payout yield: Implications for empirical asset pricing. *The Journal of Finance*, 62(2), 877-915. doi:10.2139/ssrn.480171

Buffett, M., & Clark, D. (1997). Buffettology. Rawson Associates. New York.

Campbell, J. Y., Hilscher, J., & Szilagyi, J. (2008). In search of distress risk. *The Journal of Finance*, 63(6), 2899-2939. doi:10.2139/ssrn.770805

Carhart, M. M. (1997). On persistence in mutual fund performance. *The Journal of Finance*, *52*(1), 57-82. doi:10.1111/j.1540-6261.1997.tb03808.x

Casey, R. J., Gao, F., Kirschenheiter, M. T., Li, S., & Pandit, S. (2015). Do Compustat financial statement data articulate? *Journal of Financial Reporting*, 1(1), 37-59 doi:10.2139/ssrn.2380698

Cohen, R. B., Polk, C., & Vuolteenaho, T. (2003). The value spread. *The Journal of Finance*, 58(2), 609-642. doi:10.1111/1540-6261.00539

Fama, E. F., & French, K. R. (1993). Common risk factors in the returns on stocks and bonds. *Journal of Financial Economics*, 33(1), 3-56. doi:10.1016/0304-405X(93)90023-5

French, K. R. (2017) .  New York Stock Exchange market equity breakpoints [Data file]. Retrieved from http://mba.tuck.dartmouth.edu/pages/faculty/ken.french/data_library.html

Friedman, J., Hastie, T., & Tibshirani, R. (2009). *The Elements of Statistical Learning.* New York: Springer Series in Statistics. doi:10.1007/978-0-387-84858-7

Graham, B., & Dodd, D. L. (1934*). Security Analysis: Principles and Technique*. McGraw-Hill.

Graham, B. (1973). *The Intelligent Investor — fourth revised edition*.  Harper & Row

Gray, W. R., & Carlisle, T. E. (2012). *Quantitative Value: A Practitioner's Guide to Automating Intelligent Investment and Eliminating Behavioral Errors*. John Wiley & Sons.

Greenblatt, J. (2006). *The Little Book That Beats the Market*. John Wiley & Sons.

Hanley, J. A., & McNeil, B. J. (1982). The meaning and use of the area under a receiver operating characteristic (ROC) curve. *Radiology*, 143(1), 29-36. doi:10.1148/radiology.143.1.7063747

James, G., Witten, D., Hastie, T., & Tibshirani, R. (2013). *An Introduction to Statistical Learning* (Vol. 6). New York: Springer. doi:10.1007/978-1-4614-7138-7

Koh, P. S., & Reeb, D. M. (2015). Missing R&D. *Journal of Accounting and Economics*, 60(1), doi:73-94. 10.1016/j.jacceco.2015.03.004

Kok, U. W., Ribando, J., & Sloan, R. (2017). Facts about Formulaic Value Investing. *Financial Analysts Journal*, *73*(2), 81-99. doi:10.2469/faj.v73.n2.2

Lee, C. M. (2014). Value investing: Bridging theory and practice. *China Accounting and Finance Review*, *16*(2), 1-29. doi:0.7603/s40570-014-0005-3

McDonald, J. H. (2009). *Handbook of Biological Statistics* (Vol. 2, pp. 173-181). Baltimore, MD: Sparky House Publishing.

Officer, R. R. (1972). The distribution of stock returns. *Journal of the American Statistical Association*, 67(340), 807-812. doi:10.1080/01621459.1972.10481297

Piotroski, J. D. (2000). Value investing: The use of historical financial statement information to separate winners from losers. *Journal of Accounting Research*, 1-41. doi:10.2307/2672906

Sloan, R. (1996). Do stock prices fully reflect information in accruals and cash flows about future earnings? *Accounting Review*, 71(3), 289-315. doi:10.2469/dig.v27.n1.5

Wilcoxon, F. (1945). Individual comparisons by ranking methods. *Biometrics Bulletin*, *1*(6), 80-83. doi:10.2307/3001968